\documentclass[draftmode]{AEA}
\usepackage{natbib}

\usepackage{graphicx}
\usepackage{hyperref}
\usepackage{booktabs}

\usepackage{amsmath}
\usepackage{amssymb}
\usepackage{afterpage}

\usepackage[colorinlistoftodos,prependcaption,textsize=small,backgroundcolor=yellow,linecolor=lightgray,bordercolor=lightgray]{todonotes}
\newcommand{\figuredir}{.}

\begin{document}

\title{Estimating Heterogeneous Consumer Preferences for Restaurants and Travel Time Using Mobile Location Data}

\shortTitle{Consumer Preferences for Restaurants}

\author{Susan Athey, David Blei, Robert Donnelly, Francisco Ruiz and
  Tobias Schmidt\thanks{Athey: Stanford University, 655 Knight Way,
    Stanford, CA 94305, athey@stanford.edu.  Blei: Columbia
    University, Department of Computer Science, New York, NY, 10027,
    david.blei@columbia.edu. Donnelly: Stanford University, 655 Knight
    Way, Stanford, CA 94305, rodonn@stanford.edu. Ruiz: Columbia
    University, Department of Computer Science, New York, NY, 10027,
    fr2392@columbia.edu, and University of Cambridge, Department of
    Engineering, Cambridge CB2 1PZ, UK. Schmidt: Stanford University,
    655 Knight Way, Stanford, CA 94305, tobiass@stanford.edu. The authors are listed in alphabetical order. We are
    grateful to SafeGraph and Yelp for providing the data, and to
    Paula Gablenz, Renee Reynolds, Tony Fan, and Arjun Parthipan for
    exceptional research assistance.  We acknowledge generous
    financial support from Microsoft Corporation, the Sloan
    Foundation, the Cyber Initiative at Stanford, and the Office of
    Naval Research. Ruiz is supported by the EU H2020 programme (Marie
    Sk\l{}odowska-Curie grant agreement 706760).}}\pubMonth{May}
\date{\today} \pubYear{2018} \pubVolume{Vol} \pubIssue{Issue}
\begin{abstract}
\small
This paper analyzes consumer choices over lunchtime restaurants using data from a sample of several thousand anonymous mobile phone users in the San Francisco Bay Area. The data is used to identify users’ approximate typical morning location, as well as their choices of lunchtime restaurants. We build a model where restaurants have latent characteristics (whose distribution may depend on restaurant observables, such as star ratings, food category, and price range), each user has preferences for these latent characteristics, and these preferences are heterogeneous across users. Similarly, each item has latent characteristics that describe users’ willingness to travel to the restaurant, and each user has individual-specific preferences for those latent characteristics. Thus, both users’ willingness to travel and their base utility for each restaurant vary across user-restaurant pairs. We use a Bayesian approach to estimation. To make the estimation computationally feasible, we rely on variational inference to approximate the posterior distribution, as well as stochastic gradient descent as a computational approach. Our model performs better than more standard competing models such as multinomial logit and nested logit models, in part due to the personalization of the estimates. We analyze how consumers re-allocate their demand after a restaurant closes to nearby restaurants versus more distant restaurants with similar characteristics, and we compare our predictions to actual outcomes. Finally, we show how the model can be used to analyze counterfactual questions such as what type of restaurant would attract the most consumers in a given location.
\end{abstract}
\maketitle

%\listoftodos

Where should a a new restaurant be located?  What type of restaurant
would be best in a given location? How close does a competitor need to
be to matter?  These are examples of questions about product
design and product choice.  While there is extensive literature on
consumer response to prices, there is relatively little attention to firm
choices about physical location and product characteristics.  Recent trends
in digitization have led to the creation of many
large panel datasets of consumers, which in turn motivates the development
of models that exploit the rich information in the data and provide precise
answers to these questions.

Answering many of these questions requires a model that incorporates individual-level heterogeneity in
preferences for product attributes and travel time, as these
characteristics might vary substantially even within a city.  More
broadly, understanding individual heterogeneity in travel preferences
is a key input for urban planning.  To this end, we develop an
empirical model of consumer choices over lunchtime restaurants, the
Travel-Time Factorization Model (TTFM).  TTFM incorporates rich
heterogeneity in user preferences for both observed and unobserved
restaurant characteristics as well as for travel time.  We apply the
model to a dataset derived from mobile phone locations for several
thousand anonymized mobile phone users in the San Francisco Bay Area; this is the
first structural model of individual travel choice based on
mobile location data.

TTFM can answer counterfactual questions.  For example, what would
happen if a restaurant with a given set of characteristics opened or
closed in a particular location?  Using data about several hundred
openings and closings of restaurants, we compare TTFM's predictions to
the real outcomes. TTFM can also make personalized
predictions for individuals and restaurants.  Its personalized predictions are more
accurate than existing methods, especially for high-activity
individuals and popular restaurants.

TTFM incorporates recently developed approaches from machine learning
for estimating models with a large number of latent variables.  It
uses a standard discrete choice framework to model each user's choice
over restaurants, inferring the parameters of the users' utility
functions from their choice behavior. TTFM differs from more
traditional models in the number of latent variables; it incorporates
a vector of latent characteristics for each restaurant as well as
latent user preferences for these characteristics.  In addition, it
incorporates heterogeneous user preferences for travel distance, which
vary by restaurant.  These distance preferences are represented as the
inner product of restaurant-specific factors and user willingness to
travel to restaurants with those factors.  Finally, TTFM is a
hierarchical model, where observable restaurant characteristics affect
the distribution of latent restaurant characteristics.  We use a
Bayesian approach to inference, where we estimate posterior
distributions over each user's preferences and each restaurant's
characteristics. The posterior is complex and the dataset is large.
Thus, to make the estimation computationally feasible, we rely on
stochastic variational inference to approximate the posterior
distribution with a stochastic gradient optimization algorithm.

Our approach builds on a large literature in economics and marketing
on estimating discrete choice models of consumer behavior; see
\citet{keane2013panel} for a survey. It also relates to a decades-old
literature in marketing on inferring ``product maps'' from panel data
\citep{elrod1988choice}.  Our estimation strategy is drawn from
approaches developed in \citet{athey2017counterfactual} and
\citet{ruiz2017shopper}, both of which considered the problem of
choosing items from a supermarket, and it also relates to
\citet{Wan2017}, who take a matrix factorization approach to consumer
choice.  Though less well-studied, there has also been some work on
estimating consumer preferences for travel time, e.g.,
\citet{neilsontargeted}'s study of school choice.

\section{Empirical Model and Estimation}

We model the consumer's choice of restaurant conditional on deciding
to go out to lunch.  We assume that the consumer selects the
restaurant that maximizes utility, where the utility of user $u$ for
restaurant $i$ on her $t$-th visit is
\[U_{uit} = \lambda_i + \theta_u^\top \alpha_i + \mu_i^\top \delta_{w_{ut}} - \gamma_u^\top \beta_i \cdot \text{log}(d_{ui}) + \epsilon_{uit},
\]
where $w_{ut}$ denotes the week in which trip $t$ happens, and
$d_{ui}$ is the distance from $u$ to $i$. This gives a parameterized
expression for the utility: $\lambda_i$ is an intercept term that
captures a restaurant's popularity; $\theta_u$ and $\alpha_i$ are
latent vectors that model a user's latent preferences and a
restaurant's latent attributes; $\beta_i$ is a vector that captures a
restaurant's latent factors for travel distance and $\gamma_u$ is a
user's latent preferences of willingness to travel to restaurants with
those factors; $\delta_w$ and $\mu_i$ are latent vectors of
week/restaurant time effects (this allows us to capture varying
effects for different parts of the year); and $\epsilon_{uit}$ are
error terms, which we assume to be independent and identically Gumbel
distributed.  We specify a hierarchical model where observable
characteristics of restaurants, denoted by $x_i$, affect the mean of
the distribution of latent restaurant characteristics $\alpha_i$ and
$\beta_i$. This hierarchy allows restaurants to share statistical
strength, which helps to infer the latent variables of low-frequency
restaurants. We estimate the posterior over the latent model
parameters using variational inference. Our approach is similar to
\citet{ruiz2017shopper}, but differs in a few respects.  First, we
assume that each consumer chooses only one restaurant on a purchase
occasion, so interactions among products are not important.  Second,
TTFM is hierarchical, allowing observed restaurant characteristics to
affect the prior distribution of latent variables.  (See
Appendix~\ref{sec:estdetails} for details.)

For comparison, we also consider a simpler model, a standard
multinomial logit model (MNL), which is a restricted version of our
proposed model: the term $\lambda_i$ is constant across restaurants,
$\alpha_i$ is set to be equal to the observable characteristics of
items, $\theta_u$ is constant across users, $\delta_w$ is omitted (including it created problems with convergence
of the estimation),
and $\gamma_u \cdot \beta_i$ is restricted to be constant across users
and restaurants.

\section{The Data and Summary Statistics}

The dataset is from SafeGraph, a company that collects anonymous, aggregates locational
information from consumers who have opted into sharing their location
through mobile applications.  The data consists of ``pings'' from
consumer phones; each observation includes a unique device identifier that we
associate with a single anonymous consumer, the time and date of the ping, and
the latitude, longitude and accuracy of the ping over a sample period from January
through October 2017.
% sample period checked by TS on 01/16

From this data, we construct the key variables for our
analysis. First, we construct the approximate ``typical'' morning location of the
consumer, defined as the most common place the consumer is found from
9:00 to 11:15 a.m.\ on weekdays.  We restrict attention to consumers
whose morning locations are consistent over the sample period, and for
which these locations are in the Peninsula of the San Francisco Bay
Area (roughly, South San Francisco to San Jos\'{e}, excluding the
mountains and coast). We determine that the consumer visited a
restaurant for lunch if we observed at least two pings more than 3
minutes apart during the hours of 11:30 a.m.\ to 1:30 p.m.\ in a
location that we identify as a restaurant. Restaurants are identified
using data from Yelp that includes geo-coordinates, star ratings,
price range, restaurant categories (e.g., Pizza or Chinese), and we
also use Yelp to infer approximate dates of restaurant openings and
closings.  Last, we narrow the dataset to consumer choices over a
subset of restaurants that appear sufficiently often in the data, and
to consumers who visit a sufficient number of restaurants.  This
process results in a final dataset of 106,889 lunch visits by 9,188
users to 4,924 locations.
% time periods checked by TS on 01/16
% number checked by TS on 01/14
% generated by source/analysis/descriptives/raw_data_descriptives.R
Table~\ref{tab:summary} provides summary
statistics on the users and restaurants included in the dataset.
(Appendix~\ref{sec:sampleselect} gives all details about the dataset
processing pipeline.)

\begin{table}
  \caption{Summary Statistics.\label{tab:summary}}
  \begin{tabular}{lccccc}
  \hline 
 \multicolumn{6}{c}{User-Level Statistics} \\ 
 \hline 
 Variable (Per User) & Mean & 25\% & 50\% & 75\% & \% Missing \\ 
 \hline 
 Total Visits & 11.63 & 4.00 & 7.00 & 13.00 & --- \\ 
  Distinct Visited Rest. & 7.25 & 3.00 & 5.00 & 9.00 & --- \\ 
  Distinct Visited Categories & 11.60 & 6.00 & 10.00 & 15.00 & --- \\ 
  Median Distance (mi.) & 3.06 & 0.89 & 1.86 & 3.79 & --- \\ 
  Weekly Visits & 0.39 & 0.15 & 0.25 & 0.47 & --- \\ 
  Weeks Active & 31.14 & 22.00 & 33.00 & 41.00 & --- \\ 
  Mean Rating of Visited Rest. & 3.29 & 3.00 & 3.33 & 3.61 & 1 \\ 
  Mean Price Range of Visited Rest. & 1.55 & 1.33 & 1.53 & 1.75 & 0.6 \\ 
   \hline 
 \multicolumn{6}{c}{Restaurant-Level Statistics} \\ 
 \hline 
 Variable (Per Restaurant) & Mean & 25\% & 50\% & 75\% & \% Missing \\ 
 \hline 
Distinct Visitors & 13.53 & 5.00 & 10.00 & 19.00 & --- \\ 
  Median Distance (mi.) & 2.39 & 0.93 & 1.72 & 2.94 & --- \\ 
  Weeks Open & 42.17 & 44.00 & 44.00 & 44.00 & --- \\ 
  Weekly Visits (Opens) & 0.54 & 0.17 & 0.37 & 0.72 & --- \\ 
  Weekly Visits (Always Open) & 0.52 & 0.16 & 0.34 & 0.68 & --- \\ 
  Weekly Visits (Closes) & 0.53 & 0.15 & 0.34 & 0.67 & --- \\ 
  Price Range & 1.56 & 1.00 & 2.00 & 2.00 & 10.66 \\ 
  Rating & 3.38 & 2.89 & 3.53 & 4.00 & 14.52 \\ 
   \bottomrule
\end{tabular}

\end{table}
\afterpage{\clearpage}
\section{Estimation and Model Fit}

We divide the dataset into three parts, 70.6 percent training, 5.0 percent
validation, and 24.4 percent testing.
% numbers checked by TS on 01/14
% generated by source/analysis/descriptives/raw_data_descriptives.R
We use the validation dataset to
select parameters such as the length of the latent vectors $\alpha_i$
and $\beta_i$ ($k_1$ and $k_2$, respectively), while we compare models
and evaluate performance in the test dataset.  (See
Section~\ref{sec:fit} for details.) We
select $k_1 = 80$ and $k_2 = 16$.  In the hierarchical prior,
the distribution of a restaurant's components depends on price range,
star ratings, and restaurant
category.

Across several measures evaluated on the test set, TTFM is a better model than
MNL.  For example, precision@5 is the percentage of times that a user's
chosen restaurant is in the set of the top five predicted restaurants.
It is 35\% for TFMM and 11\% for MNL. Further, as shown in
Figures~\ref{fig:fitbyuserdecile} and~\ref{fig:fitbyrestdecile}, TTFM
predictions improve significantly for high-frequency users and
restaurants, while MNL does not exhibit that improvement.  This
highlights the benefits of personalization: When given enough data,
TTFM learns user-specific preferences.

\begin{figure}
  \begin{minipage}{\linewidth} \centering
    \includegraphics[width = \textwidth]{\figuredir/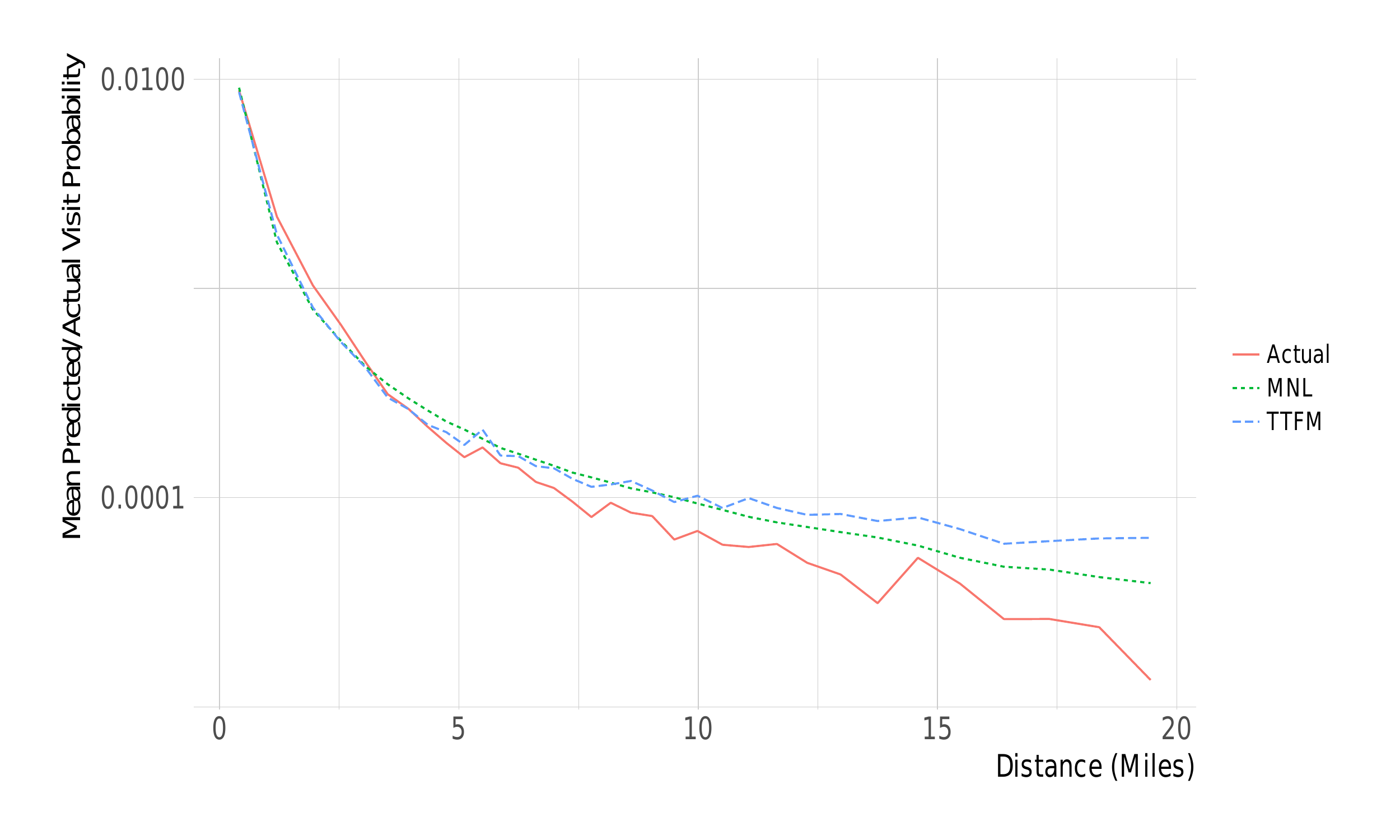}
    \caption{Predicted Versus Actual Shares By Distance}\label{fig:demandbydistance}
  \end{minipage}
  \begin{minipage}{\linewidth} \centering
    \includegraphics[width = \textwidth]{\figuredir/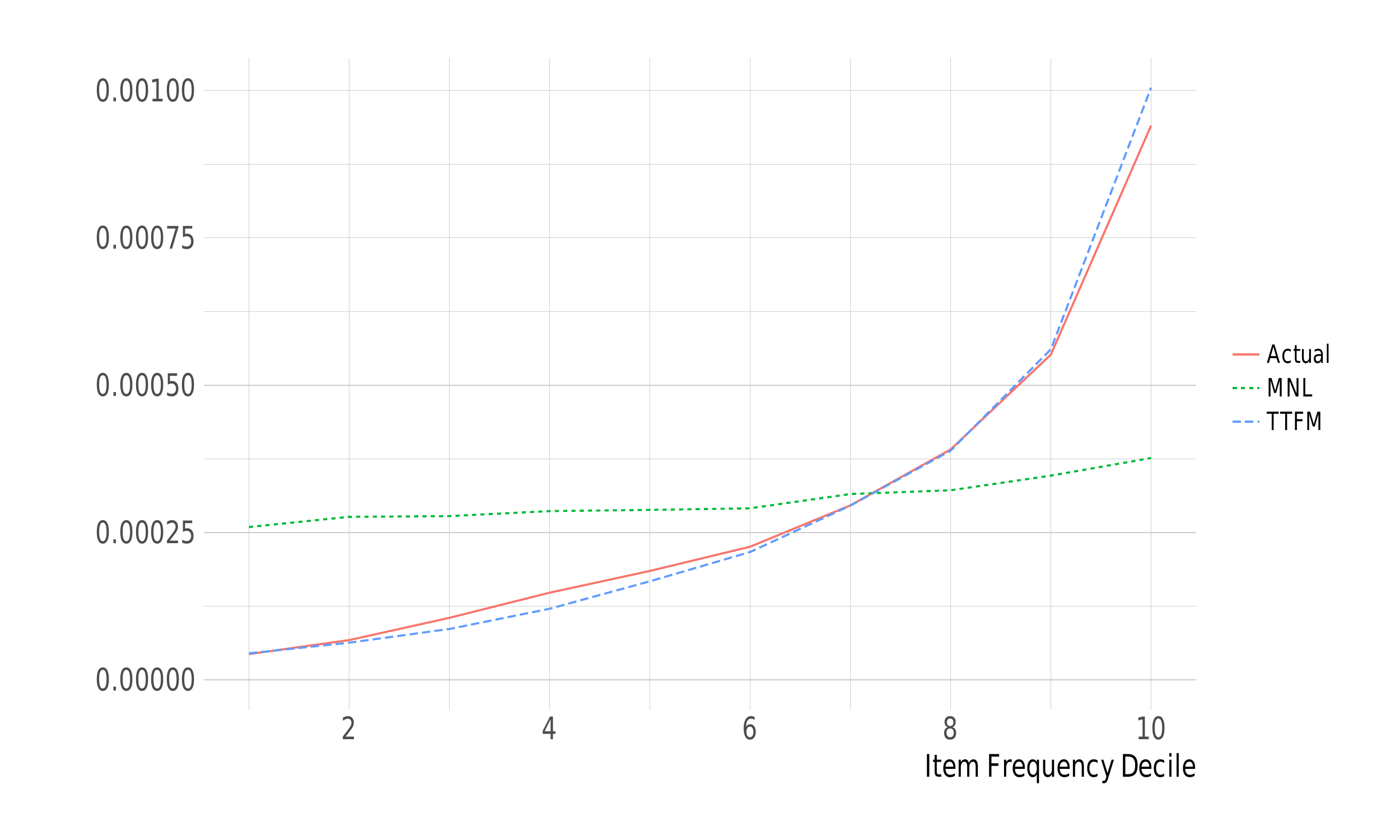}
    \caption{Predicted Versus Actual Shares by Restaurant Visit Decile}\label{fig:predvactual-itemdec}
  \end{minipage}
\end{figure}

\begin{table}[!h]
  \centering
  \caption{Goodness of Fit of Alternative Models}\label{fig:modelfitsummary}
  \centering
\resizebox{\linewidth}{!}{\begin{tabular}{llllll}
\toprule
Model & MSE & Log Likelihood & Precision@1 & Precision@5 & Precision@10\\
\midrule
\addlinespace[0.3em]
\multicolumn{6}{l}{\textbf{Training Sample}}\\
\hspace{1em}TTFM & 0.00025 & -3.59 & 31.8\% & 59.4\% & 70.3\%\\
\hspace{1em}MNL & 0.00031 & -6.58 & 2.8\% & 10.7\% & 16.7\%\\
\addlinespace[0.3em]
\multicolumn{6}{l}{\textbf{Held-out Test Sample}}\\
\hspace{1em}TTFM & 0.00028 & -5.19 & 20.5\% & 35.5\% & 42.2\%\\
\hspace{1em}MNL & 0.00031 & -6.55 & 3.1\% & 11.4\% & 17.5\%\\
\bottomrule
\end{tabular}}

  \begin{tablenotes}
  Precision measures the share of visits in the set of the top \{1,5,10\} restaurants predicted by the model.
  \end{tablenotes}
\end{table}
\afterpage{\clearpage}
Figure~\ref{fig:demandbydistance} illustrates that both TTFM and MNL
fit well the empirical probability of visiting restaurants at varying
distances from the consumer's morning location.  But
Figure~\ref{fig:predvactual-itemdec} shows that TTFM outperforms MNL
at fitting the actual visit rates of different restaurants; here
restaurants are grouped by their visit-frequency deciles.  The rich
heterogeneity of TTFM allows personalized predictions for restaurants.

\afterpage{\clearpage}
\section{Parameter Estimates}

The distributions of estimated elasticities from TTFM are summarized in Table \ref{tab:distance_elasticities} and
Figure~\ref{fig:elasticity}.  Note that the elasticities in the MNL
vary only because the baseline visit probabilities vary across
consumers and restaurants.  TTFM elasticities are more dispersed,
reflecting the personalization capabilities of the TTFM model.  The
average elasticity across consumers and restaurants (weighted by
trip frequency) is $-1.41$. 
% number checked by RD on 01/19
% generated by source/analysis/descriptives/distance_elasticity_descriptives.R
% output in release/tables/distance/summary_stats_distance_elasticities.tex
Thus, distance matters substantially for lunch, which is consistent with the fact that
roughly 60 percent of visits are within two miles of the consumer's
morning location.
% number checked by TS on 01/13
% generated by source/analysis/descriptives/raw_data_descriptives.R
% 56.96488 % of trips go to locations less than 2 miles from the morning location
% 68.75166 % of trips go to locations less than 3 miles from the morning location
% 58.51409 % of trips below 20km go to locations less than 2 miles from the morning location
% 70.62142 % of trips below 20km go to locations less than 3 miles from the morning location
Furthermore, there is substantial heterogeneity in that willingness to
travel. Across users and restaurants, the standard deviation of elasticities in the TTFM model is 0.68, while the average within-user standard deviation of elasticities is 0.30 and the average within-restaurant standard deviation of elasticities is 0.60. Elasticities are substantially less dispersed in the MNL model.
% all numbers checked by RD on 01/19
% generated by source/analysis/descriptives/distance_elasticity_descriptives.R
% output in release/tables/distance/summary_stats_distance_elasticities.tex

\begin{table}
  \caption{Average Within-Item Elasticities by Restaurant Characteristics, TTFM model.\label{tab:prefdistancebytype_ttfm}}
  \begin{tabular}{lrrrrrr}
  \toprule
Characteristic & Mean & se & 25 \% & 50 \% & 75 \% & N \\ 
  \midrule
All restaurants & -1.411 & 0.0001 & -1.585 & -1.408 & -1.203 &  4924 \\ 
  Most popular category: Mexican & -1.499 & 0.0004 & -1.664 & -1.491 & -1.285 &   694 \\ 
  Most popular category: Sandwiches & -1.435 & 0.0006 & -1.602 & -1.441 & -1.235 &   522 \\ 
  Most popular category: Hotdog & -1.403 & 0.0007 & -1.570 & -1.390 & -1.216 &   377 \\ 
  Most popular category: Coffee & -1.390 & 0.0008 & -1.563 & -1.404 & -1.178 &   365 \\ 
  Most popular category: Bars & -1.370 & 0.0009 & -1.546 & -1.362 & -1.161 &   352 \\ 
  Most popular category: Chinese & -1.353 & 0.0009 & -1.517 & -1.378 & -1.176 &   350 \\ 
  Most popular category: Japanese & -1.320 & 0.0011 & -1.472 & -1.336 & -1.140 &   276 \\ 
  Most popular category: Pizza & -1.497 & 0.0010 & -1.649 & -1.481 & -1.307 &   260 \\ 
  Most popular category: Newamerican & -1.323 & 0.0019 & -1.540 & -1.351 & -1.117 &   181 \\ 
  Most popular category: Vietnamese & -1.328 & 0.0020 & -1.541 & -1.327 & -1.155 &   156 \\ 
  Most popular category: Other & -1.411 & 0.0002 & -1.582 & -1.406 & -1.189 &  1391 \\ 
  Price range: 1 & -1.446 & 0.0001 & -1.607 & -1.435 & -1.245 &  2091 \\ 
  Price range: 2 & -1.368 & 0.0001 & -1.542 & -1.371 & -1.162 &  2165 \\ 
  Price range: 3 & -1.320 & 0.0026 & -1.506 & -1.353 & -1.108 &   122 \\ 
  Price range: 4 & -1.449 & 0.0178 & -1.664 & -1.496 & -1.289 &    21 \\ 
  Price range: missing & -1.474 & 0.0006 & -1.648 & -1.455 & -1.225 &   525 \\ 
  Rating, quintile: 1 & -1.427 & 0.0003 & -1.605 & -1.414 & -1.209 &   842 \\ 
  Rating, quintile: 2 & -1.392 & 0.0003 & -1.557 & -1.397 & -1.187 &   842 \\ 
  Rating, quintile: 3 & -1.364 & 0.0003 & -1.532 & -1.366 & -1.169 &   842 \\ 
  Rating, quintile: 4 & -1.385 & 0.0004 & -1.571 & -1.370 & -1.180 &   842 \\ 
  Rating, quintile: 5 & -1.438 & 0.0003 & -1.603 & -1.438 & -1.250 &   841 \\ 
  Rating, quintile: missing & -1.475 & 0.0004 & -1.653 & -1.464 & -1.232 &   715 \\ 
   \bottomrule
\end{tabular}

\end{table}

\begin{table}
  \caption{Average Within-Item Elasticities by City, TTFM model.}\label{tab:prefdistancebycity_ttfm}
  \begin{tabular}{lrrrrrr}
  \toprule
Characteristic & Mean & se & 25 \% & 50 \% & 75 \% & N \\ 
  \midrule
All restaurants & -1.411 & 0.0001 & -1.585 & -1.408 & -1.203 &  4924 \\ 
  City: Daly City & -1.105 & 0.0019 & -1.331 & -1.150 & -0.959 &   165 \\ 
  City: Burlingame & -1.119 & 0.0030 & -1.327 & -1.194 & -1.018 &   110 \\ 
  City: Millbrae & -1.130 & 0.0049 & -1.418 & -1.240 & -0.954 &    80 \\ 
  City: San Bruno & -1.132 & 0.0035 & -1.398 & -1.216 & -0.987 &   101 \\ 
  City: South San Francisco & -1.187 & 0.0021 & -1.413 & -1.232 & -0.999 &   135 \\ 
  City: San Mateo & -1.243 & 0.0012 & -1.454 & -1.284 & -1.101 &   268 \\ 
  City: Foster City & -1.318 & 0.0070 & -1.506 & -1.397 & -1.163 &    44 \\ 
  City: San Carlos & -1.321 & 0.0026 & -1.479 & -1.350 & -1.195 &    95 \\ 
  City: Palo Alto & -1.330 & 0.0013 & -1.519 & -1.342 & -1.171 &   234 \\ 
  City: Brisbane & -1.332 & 0.0139 & -1.455 & -1.344 & -1.181 &    15 \\ 
  City: Belmont & -1.334 & 0.0047 & -1.500 & -1.374 & -1.212 &    58 \\ 
  City: Redwood City & -1.362 & 0.0012 & -1.530 & -1.389 & -1.217 &   214 \\ 
  City: Cupertino & -1.365 & 0.0018 & -1.532 & -1.386 & -1.174 &   169 \\ 
  City: East Palo Alto & -1.374 & 0.0142 & -1.521 & -1.393 & -1.229 &    13 \\ 
  City: Los Gatos & -1.391 & 0.0026 & -1.583 & -1.437 & -1.219 &   106 \\ 
  City: Los Altos & -1.406 & 0.0043 & -1.564 & -1.394 & -1.236 &    60 \\ 
  City: Menlo Park & -1.407 & 0.0031 & -1.570 & -1.428 & -1.287 &    87 \\ 
  City: Mountain View & -1.422 & 0.0013 & -1.592 & -1.429 & -1.233 &   213 \\ 
  City: Santa Clara & -1.442 & 0.0009 & -1.681 & -1.456 & -1.238 &   355 \\ 
  City: San Jose & -1.451 & 0.0002 & -1.635 & -1.464 & -1.278 &  1858 \\ 
  City: Campbell & -1.482 & 0.0015 & -1.640 & -1.493 & -1.317 &   144 \\ 
  City: Saratoga & -1.497 & 0.0059 & -1.628 & -1.481 & -1.394 &    40 \\ 
  City: Sunnyvale & -1.501 & 0.0008 & -1.659 & -1.513 & -1.325 &   302 \\ 
  City: Stanford & -1.607 & 0.0062 & -1.760 & -1.605 & -1.482 &    39 \\ 
   \bottomrule
\end{tabular}

\end{table}

\begin{figure}
  \includegraphics[width=.8 \textwidth]{\figuredir/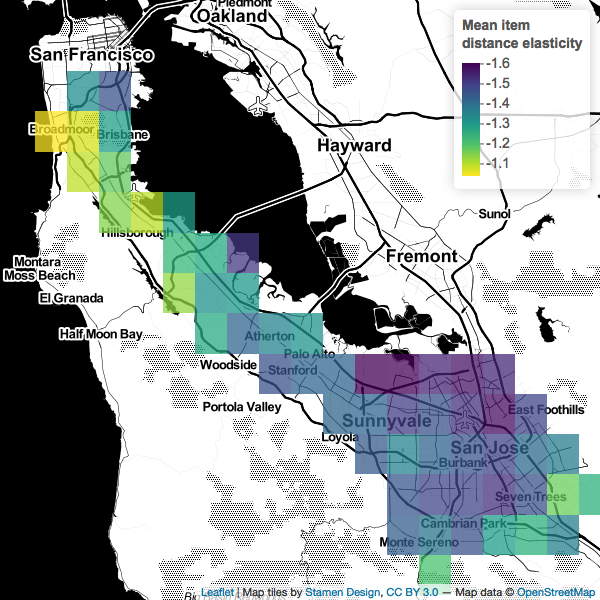}
  \caption{Average Within-Item Elasticities by geohash6, TTFM model.} \label{fig:elasticityprefbygeo}
\end{figure}
\afterpage{\clearpage}
Tables~\ref{tab:prefdistancebytype_ttfm} and~\ref{tab:prefdistancebycity_ttfm} and Figure \ref{fig:elasticityprefbygeo} illustrate how elasticities vary across restaurant types and cities. Willingness to travel
is lower for low-priced restaurants (elasticity $-1.45$
for price range \$ (under \$10) versus $-1.37$ for price range \$\$
(\$11$-$\$30)); lower
for Mexican restaurants and Pizza places than for Chinese and Japanese
restaurants (elasticities of $-1.50$ and $-1.50$ versus $-1.35$ and $-1.32$,
respectively). Cities with many work locations nearby retail districts, including San Jos\'{e}, Sunnyvale, and Mountain
View have a lower willingness to travel than cities that are more spread out like Daly City, Burlingame,
San Bruno, and San Mateo.  Appendix
Section~\ref{sec:addlresults} provides further descriptive statistics
about latent factors and model results, illustrating for example how to model can be used to find restaurants that are intrinsically similar
(without regard to location) as well as which restaurants are similar in terms of user utilities.
% all numbers checked by RD on 01/19
% generated by source/analysis/descriptives/distance_elasticity_descriptives.R
% output in release/tables/distance/item_distance_elasticities_by_city_mnl.tex and item_distance_elasticities_by_other_covariates_mnl.tex, both included in appendix

\section{Analyzing Restaurant Opening and Closing}\label{sec:openclose}

The TTFM model can make predictions about how market share will be redistributed among restaurants when restaurants open or close, and these predictions can be compared to the actual changes that occur in practice.  For this exercise, we focus on 221 openings and 190 closings where, both before and after the change, there
were at least 500 restaurant visits by users with morning locations within a 3 mile radius of the relevant restaurant.  Figure~\ref{fig:openclose} illustrates that restaurant openings and closings are fairly evenly distributed over the time period.
% numbers checked by TS on 01/14
% events generated by source/analysis/substitution_effects/openings_and_closings_analyze.R

One challenge of analyzing market share redistribution is that for any given target restaurant that opens or closes, we would expect some baseline level of market share changes of competing restaurants due to changes in the open status of neighboring restaurants.  We address this in an initial exercise where we hold the environment fixed in the following way. For each target restaurant that changed status, we first construct the predicted difference in market shares for each other restaurant between the ``closed'' and ``open'' regime (irrespective of which came first in time), and then subtract out the predicted change in market share that would have occurred for each restaurant if the target restaurant had been closed in both periods. We then sum the changes across restaurants in different groups defined by their distance from the target restaurant.   Table~\ref{tab:opcl_distance_redistribution_shares} shows TTFM model predictions for how the opening/closing restaurant's market share is redistributed over other restaurants within certain distances after the restaurant becomes unavailable (i.e.~before the opening or after the closing).  The TTFM model estimates imply that just over 50 percent of the market share impact of a closure accrues restaurants within 2 miles of the target restaurant.

\begin{table}
  \caption{Share of demand redistributed by distance, TTFM model relative to benchmark}\label{tab:opcl_distance_redistribution_shares}
  \begin{tabular}{lrrrrrr}
  \toprule  & \multicolumn{6}{c}{Distance from opening/closing restaurant (mi.)}\\ \cmidrule(lr){2-7}  & $<$ 2  & 2 - 4  & 4 - 6  & 6 - 8  & 8 - 10  & $>$ 10  \\ 
  \midrule share & 51 \% & 23 \% & 10 \% & 6 \% & 3 \% & 6 \% \\ 
  cum. share & 51 \% & 74 \% & 84 \% & 90 \% & 94 \% & 100 \% \\ 
   \bottomrule \end{tabular}

\end{table}

% number checked by TS on 01/16
% generated by source/analysis/substitution_effects/openings_and_closings_analyze.R
% output in release/tables/distance/opcl_distance_redistribution_shares.tex included in the appendix

\begin{figure}
  \includegraphics[width=.8 \textwidth]{\figuredir/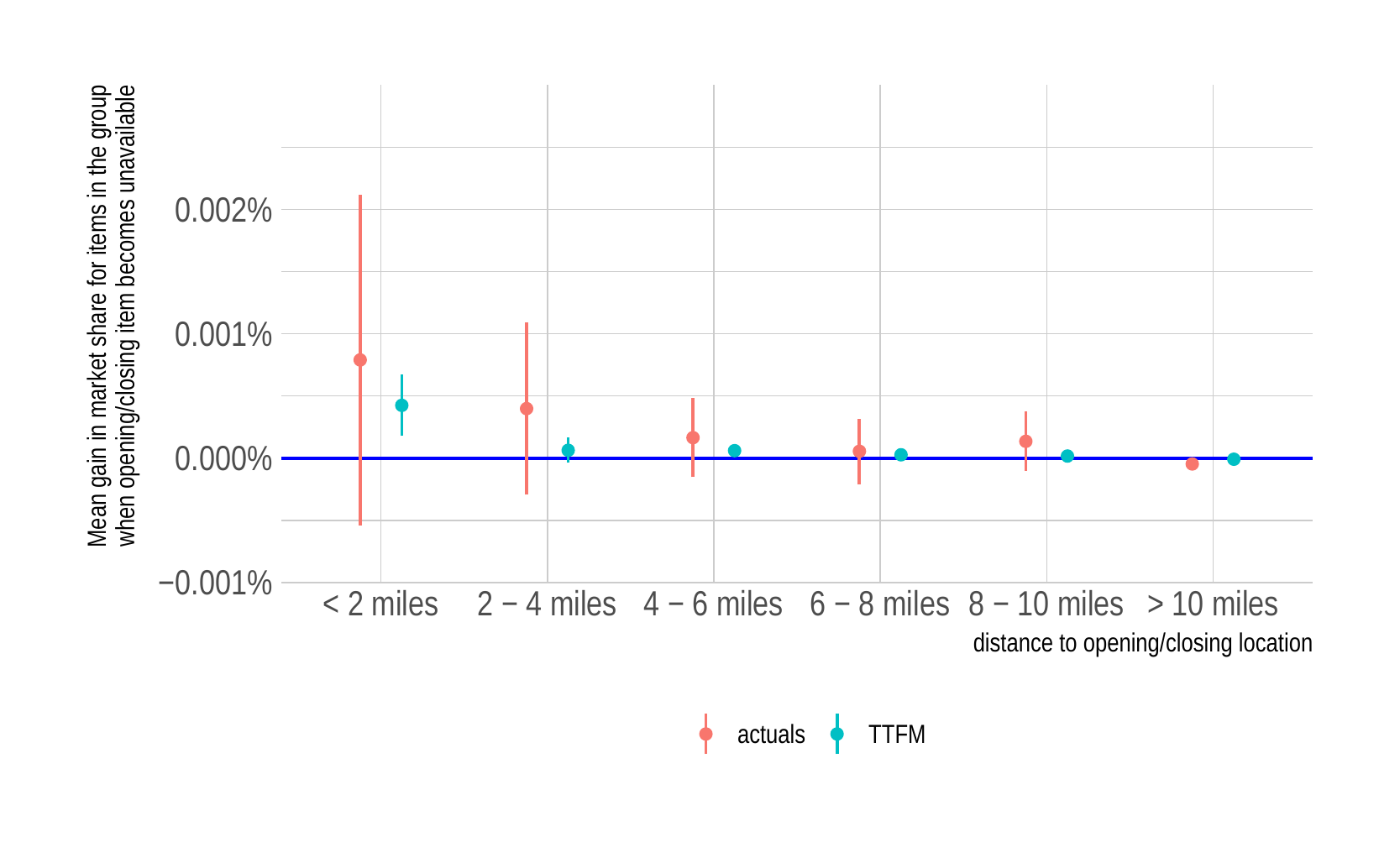}
  \caption{Model Predictions Compared to Actual Outcomes for Restaurant Openings and Closings.}\label{fig:OPCL2}
  \begin{figurenotes}
  The figure shows the average of the predicted difference in the market share of each restaurant in the group between the period where the target restaurant is closed and when it is open. The user base for the calculated market shares includes all users whose morning location is within three miles of the target restaurant and who visit at least one restaurant in both periods. We consider only restaurants that appear in the consideration sets of these users at least 500 times in both periods. User-item market shares under each regime (target restaurant open and target restaurant closed) are averaged using weights proportional to each user's share of visits in the group to any location during the open period. The bars in the figure show the point estimates plus or minus two times the standard error of the estimate, which is calculated as the standard deviation of the estimates across the different opening or closing events divided by the square root of the number of events.
  \end{figurenotes}
\end{figure}

Figure~\ref{fig:OPCL2} compares the actual changes in market share that occured against the predictions of the TTFM model.  It should be noted that baseline changes unrelated to the opening and closing of the target restaurants seem to dominate both the actual and predicted market share changes in the figure.  The figure shows that our model's predictions match well the actual changes that occurred, but it there is substantial variation in the changes that occured in the actual data, making it difficult to evaluate model performance using this exercise.

Our final exercise considers the best choice of restaurant type for a location.
For the set of restaurants that open or close, we look at how the demand for the restaurant
that changed status (the ``target restaurant'') compares to the
counterfactual demand the model predicts in the scenario where a different restaurant
in our sample (as described by its mean latent characteristics)
is placed in the location of the target restaurant.
For each target, we consider a set of 200 alternative restaurants, 100 from the
same category as the target restaurant and 100 from a different category.%
\footnote{These alternatives are sampled with equal probabilities from the set of restaurants in our
  sample.} 
We then compare the target restaurant's estimated market share to
the mean demand across the set of alternatives.
In Table~\ref{tab:CF}, we see that both the restaurants that opened and those
that closed on average have higher predicted demand than either group of
alternatives. However, the restaurants that opened appear to be in more
valuable locations, since for the 200 alternative
restaurants, we predict higher average demand if they were (counterfactually) placed
at the opening locations than at the locations of closing restaurants. As a
further comparison, we split the set of alternatives into groups based on whether or not
they are in the same broad category as the restaurant that opened or closed.
We find that alternative restaurants from the same category as the target would
perform better on average than alternatives from a different category.

\begin{table}
  \caption{Alternative Restaurant Characteristics for Opening and Closing Restaurants}\label{tab:CF}
  
\begin{tabular}{lll}
\toprule
Mean Predicted Demand & Closing & Opening\\
\midrule
Actual Opening/Closing Restaurant & 10.33 (0.83) & 12.10 (1.14)\\
Alternative from Same Category & 10.08 (0.12) & 10.53 (0.11)\\
Alternative from Different Category & 9.09 (0.08) & 9.71 (0.08)\\
\bottomrule
\end{tabular}
\end{table}

\section{Ideal Locations and Ideal Restaurant Types}\label{sec:marketfit}
In this section, we consider the match between restaurant characteristics and locations.
In each geohash6, we select one restaurant location at
random and use the TTFM model to predict what the total demand would have been
if a different restaurant had been located in its place. The set of alternative
restaurants was chosen to include one restaurant from each of the major
categories in the sample.\footnote{From each category, we randomly selected one
  restaurant whose market share is within $0.1$ standard deviation of the
  mean market share in the full sample.} 

In Figure \ref{fig:bestlocation}, we examine
which locations are predicted to provide the largest demand
in the lunch market for each restaurant category. We can see for example that Vietnamese restaurants are
predicted to have the highest demand in a dense region in the southeastern
portion of the map. The demand for Filipino restaurants is relatively diffuse,
whereas the demand for sandwiches is characterized by small but dense pockets of relatively high demand.

In Figure \ref{fig:bestcategory}, we group the restaurant categories into coarse groups
based on the price range and the type of cuisine.  We examine within each group which
category would have the highest total demand in each location. There is
considerable spatial heterogeneity in which restaurant category is predicted to
perform best in each location. 

\section{Conclusions}

This paper makes use of a novel dataset to analyze consumer choice: mobile location data.  We propose the TTFM model, a rich model that allows heterogeneity in user preferences for restaurant characteristics as well as for travel time, where preferences for travel time vary across restaurants as well.  We show that this model fits the data substantially better than traditional alternatives, and by incorporating recent advances in Bayesian inference, the estimation becomes tractable.  We use the model to conduct counterfactual analysis about the impact of restaurants opening and closing, as well as to evaluate how the choice of restaurant characteristics affects market share.  More broadly, we believe that with the advent of digitization, panel datasets about consumer location can be combined with rich structural models to answer questions about firm strategy as well as urban policy, and models such as TTFM can be used to accomplish these goals.
\clearpage
% Remove or comment out the next two lines if you are not using bibtex.
\bibliographystyle{aea}
\bibliography{library}
\newpage
% The appendix command is issued once, prior to all appendices, if any.
\appendix

\section{Appendix}

This Appendix begins by providing details of the data and dataset creation.  Next we provide estimation details.  Then, we provide
a variety of results about goodness of fit and our model estimates, including summaries of estimated sensitivity to distance broken
out by restaurant category and other characteristics.  Next, we provide details of our analyses of restaurant openings and closings, as well
as counterfactual analyses about the ideal locations of restaurants of different categories.

\subsection{Data Description}

Our dataset is constructed using data from SafeGraph, a company which aggregates locational information from anonymous consumers who have opted in to sharing their location through mobile applications.  The data consists of ``pings'' from consumer phones; each observation includes a unique device id that we associate with a single consumer; the time and date of the ping; and the latitude and longitude and horizontal accuracy of the ping, all for smartphones in use during the sample period from January through October 2017.

Our second data source is Yelp.  From Yelp, we obtained a list of restaurants, locations, ratings, price ranges, and categories, and we infer dates of openings and closings from the dates on which consumers created a listing on Yelp or marked a location as closed, respectively.

\subsection{Dataset Creation and Sample Selection}\label{sec:sampleselect}

Our area of interest is the corridor from South San Francisco to South San Jos\'{e} around I-101 and I-280. We start with a rough bounding box around the area, find all incorporated cities whose area intersects the bounding box and then remove Fremont, Milpitas, Hayward, Pescadero, Loma Mar, La Honda, Pacifica, Montara, Moss Beach, El Granada, Half Moon Bay, Lexington Hills and Colma from the set because they are too far from the corridor.

This leaves us with the following 41 cities: Los Gatos, Saratoga, Campbell, Cupertino, Los Altos Hills, Monte Sereno, Palo Alto, San Jos\'{e}, San Bruno, Atherton, Brisbane, East Palo Alto, Foster City, Hillsborough, Millbrae, Menlo Park, San Mateo, Portola Valley, Sunnyvale, Mountain View, Los Altos, Santa Clara, Belmont, Burlingame, Daly City, San Carlos, South San Francisco, Woodside, Redwood City, Alum Rock, Burbank, Cambrian Park, East Foothills, Emerald Lake Hills, Fruitdale, Highlands-Baywood Park, Ladera, Loyola, North Fair Oaks, Stanford and West Menlo Park.
% Number of cities checked by TS on 01/16. Is 41

\begin{figure}
  \includegraphics[width=.5 \textwidth]{\figuredir/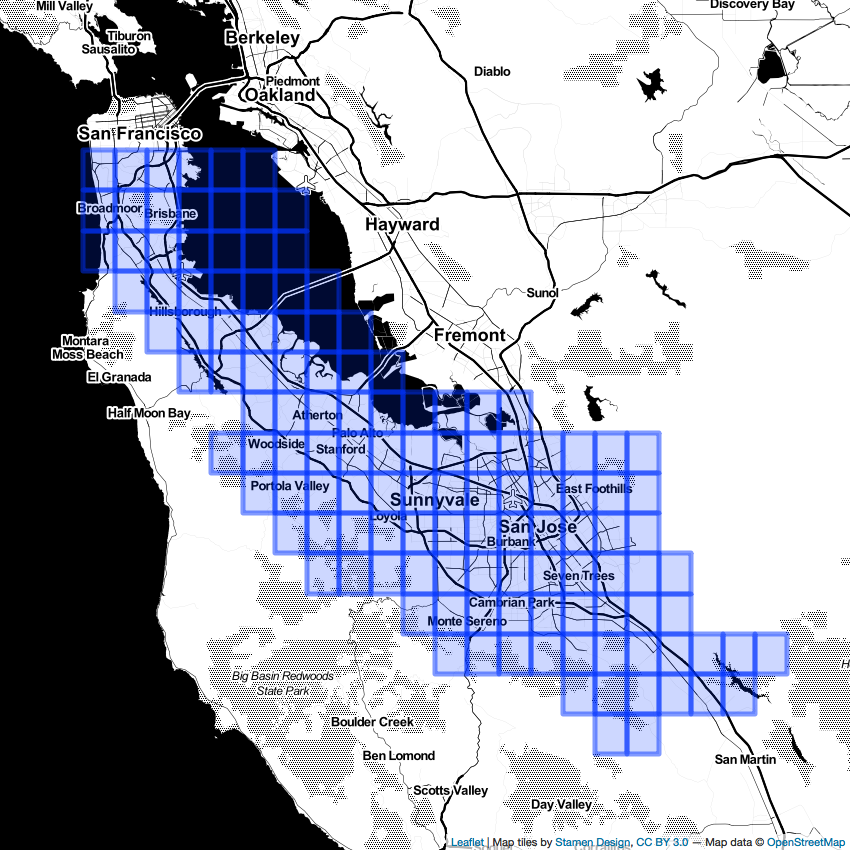}
  \caption{Geographical Region Considered}\label{fig:georegion}
\end{figure}

We then take the shapefiles for these cities as provided by the Census Bureau and find the set of rectangular regions known as geohash5s\footnote{Geohashes are a system in which the earth is gridded into a set of successively finer set of rectangles, which are then labelled with alphanumeric strings. These strings can then be used to describe geographic information in databases in a form that is easier to work with than latitudes and longitudes. At its coarsest, the geohash1 level, the earth is divided into 32 rectangles whose edges are roughly 3000 miles long. Each geohash1 is then in turn divided into 32 rectangles that are about 800 miles across. The finest geohash resolution used in this paper, geohash8, corresponds to rectangles of size 125 $\times$ 60 feet. See %Figure~\ref{fig:geohash_illustration} for an illustration and 
\url{http://www.geohash.org/} for further details.} that cover their union. This is our area of interest and is shown in Figure~\ref{fig:georegion}.
% geohash sizes checked by TS on 01/16
% Source: https://www.elastic.co/guide/en/elasticsearch/reference/current/search-aggregations-bucket-geohashgrid-aggregation.html

To construct our \emph{user base} we only consider movement pings emitted on weekdays. We define an active week to be one during which a user emits at least one such ping. The \emph{user base} includes users who meet the following criteria during our sample period, January to October 2017:

\begin{itemize}
  \setlength\itemsep{0em}
  \item Have an approximate inferred home location as provided by SafeGraph
  \item Are ``active'' (defined as having at least 12 --- not necessarily consecutive --- active weeks)
  \item Have at least 10 pings in the area of interest on average in active weeks
  \item 80 percent of pings during hours of 9 --- 11:15 a.m.~are in the area of interest
  \item 60 percent of pings during hours of 9 --- 11:15 a.m.~are in their ``broad morning location'' where ``broad morning location'' is at the geohash6 level (a rectangle of roughly 0.75 miles $\times$ 0.4 miles).
  \item 40 percent of pings during hours of 9 --- 11:15 a.m.~are in their ``narrow morning location'' where ``narrow morning location'' is at the geohash7 level (a square with edge length of roughly 500 feet).
  \item Have their ``broad morning location'' in the area of interest
 \end{itemize}

These restrictions give us 32,581 users, which we refer to as our ``user base.'' We then consider the set of restaurants.  We begin with the set of restaurants known to Yelp in the San Francisco Bay Area, which we reduce through the following restrictions:
% number checked by TS on 01/15
% generated by SELECT COUNT(1) FROM retailchoice.fct_user_base_v3 WHERE is_in_user_base = TRUE

\begin{itemize}
  \setlength\itemsep{0em}
  \item Locations are in the area of interest
  \item Locations belong not just to the category ``food'' but also belong to certain sub-categories (manually) selected from Yelp's list (\url{https://www.yelp.com/developers/documentation/v2/category_list}): thai, soup, sandwiches, juicebars, chinese, tradamerican, newamerican, bars, breweries, korean, mexican, pizza, coffee, asianfusion, indpak, delis, japanese, pubs, italian, greek, sportsbars, hotdog, burgers, donuts, bagels, spanish, basque, chicken\_wings, seafood, mediterranean, portuguese, breakfast\_brunch, sushi, taiwanese, hotdogs, mideastern, moroccan, pakistani, vegetarian, vietnamese, kosher, diners, cheese, cuban, latin, french, irish, steak, bbq, vegan, caribbean, brazilian, dimsum, soulfood, cheesesteaks, tapas, german, buffets, fishnchips, delicatessen, tex-mex, wine\_bars, african, gastropubs, ethiopian, peruvian, singaporean, malaysian, cajun, cambodian, cafes, halal, raw\_food, foodstands, filipino, british, southern, turkish, hungarian, creperies, tapasmallplates, russian, polish, afghani, argentine, belgian, fondue, brasseries, himalayan, persian, indonesian, modern\_european, kebab, irish\_pubs, mongolian, burmese, hawaiian, cocktailbars, bistros, scandinavian, ukrainian, lebanese, canteen, austrian, scottish, beergarden, arabian, sicilian, comfortfood, beergardens, poutineries, wraps, salad, cantonese, chickenshop, szechuan, puertorican, teppanyaki, dancerestaurants, tuscan, senegalese, rotisserie\_chicken, salvadoran, izakaya, czechslovakian, colombian, laos, coffeeshops, beerbar, arroceria\_paella, hotpot, catalan, laotian, food\_court, trinidadian, sardinian, cafeteria, bangladeshi, venezuelan, haitian, dominican, streetvendors, shanghainese, iberian, gelato, ramen, meatballs, armenian, slovakian, czech, falafel, japacurry, tacos, donburi, easternmexican, pueblan, uzbek, sakebars, srilankan, empanadas, syrian, cideries, waffles, nicaraguan, poke, noodles, newmexican, panasian, acaibowls, honduran, guamanian, brewpubs.\footnote{Locations can belong to several categories. The location will be included if any categories match.}
\end{itemize}

This yields a list of locations far too broad. We thus refine the resulting set of locations by removing:
\begin{itemize}
  \setlength\itemsep{0em}
  \item The coffee and tea chains Starbucks, Peet's and Philz Coffee
  \item All locations whose name matches the regular expression \texttt{(coffee|tea)} but whose name does not start with ``coffee''
  \item All locations whose name matches the regular expression \texttt{(donut|doughnut)} but does not contain ``bagel''
  \item All locations whose name matches the regular expression \texttt{food court}
  \item All locations whose name matches the regular expression \texttt{mall}
  \item All locations whose name matches the regular expression \texttt{market}
  \item All locations whose name matches the regular expression \texttt{supermarket}
  \item All locations whose name matches the regular expression \texttt{shopping center}
  \item All locations whose name matches the regular expression \texttt{(yogurt|ice cream|dessert)}
  \item All locations whose name matches the regular expression \texttt{cater} but does not match the regular expression \texttt{(and|\&)} (this is to keep places like ``Catering and Cafe'' in the sample)
  \item All locations whose name matches the regular expression \texttt{truck} and who do not have a street address (these are likely to be food trucks that move around)
  \item A number of ``false positives'' manually by name (commonly these are grocery stores, festivals or farmers' markets)
  \item A number of cafeterias at prominent Bay Area tech companies like Google, VMWare and Oracle
\end{itemize}

Finally, we review the list of locations that would be removed under these rules and save a few handsful of locations from removal manually.

\begin{figure}
  \includegraphics[width=.5 \textwidth]{\figuredir/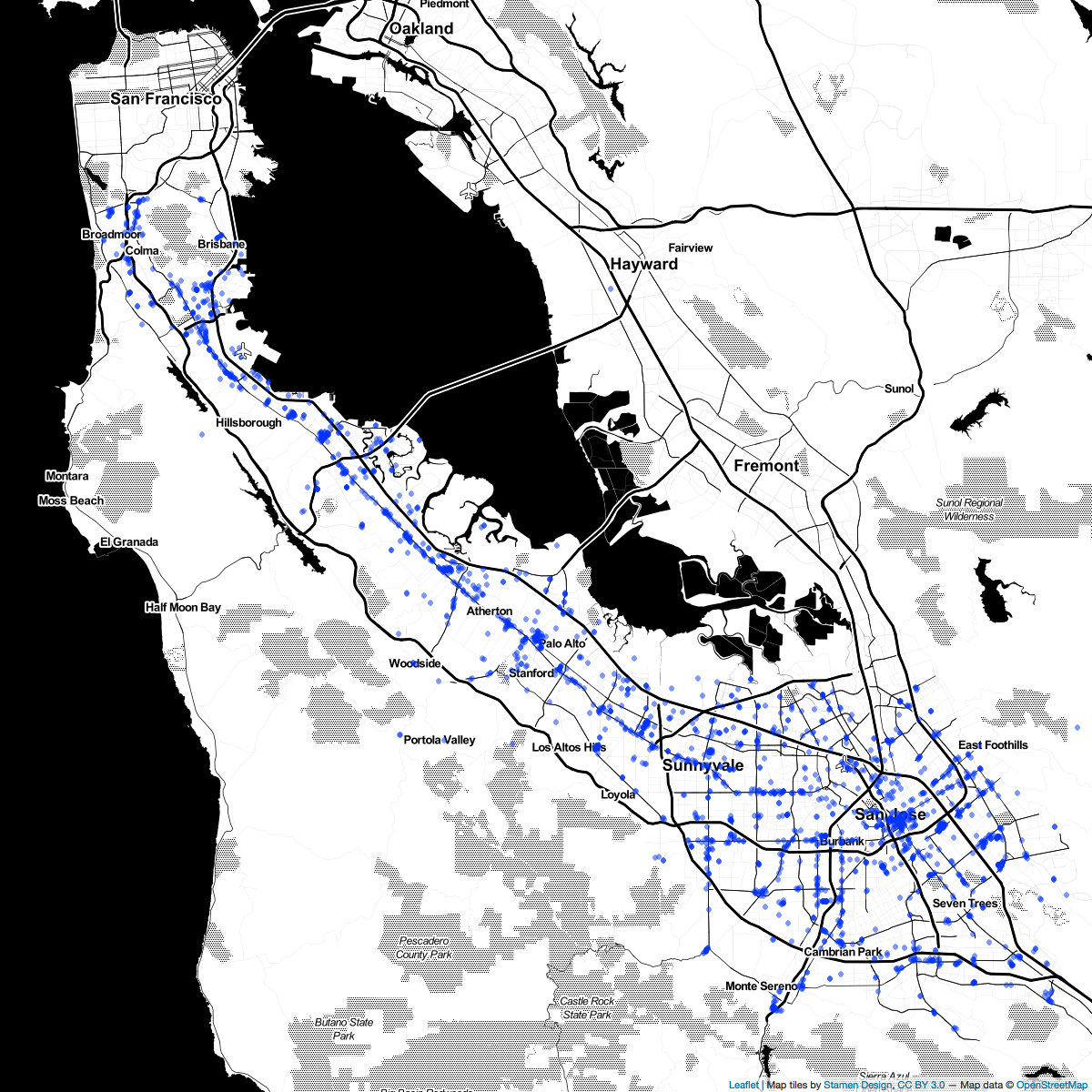}
  \caption{Included Restaurants}\label{fig:includedrestaurants}
\end{figure}

Applying these restrictions leaves us with 6,819 locations.  As a last step we de-duplicate on geohash8. Some locations are so close together that given our matching method we cannot tell them apart and need to decide which of potentially several locations in a geohash8 we want to assign a visit to. In 4,577 cases there is a unique restaurant in the geohash8, while 687 have two, with the remainder having three or more.  We de-duplicate using the first restaurant in alphabetical order, leaving us with 5,555 locations. (One reason to remove San Francisco from the sample is that higher density areas have more duplication.)  The resulting restaurants are visualized in Figure~\ref{fig:includedrestaurants}.
% number checked by TS on 01/15
% generated by 
% SELECT COUNT(1) FROM locations_v3; -> 6819
% SELECT COUNT(1) FROM locations_v3 WHERE dupe = FALSE; -> 5555
% WITH counts AS(
%     SELECT geo_hash8, COUNT(1) AS location_count
%     FROM locations_v3
%     GROUP BY geo_hash8)
% SELECT location_count, COUNT(1) AS cases FROM counts GROUP BY location_count ORDER BY cases DESC; -> 4577, 687

\afterpage{\clearpage}

Next, we define a ``visit'' to a restaurant.  For each user, each restaurant and each day we count the number of pings in the restaurant's geohash8 as well as its immediately adjacent geohash8s as well as the dwelltime, defined as the difference between the earliest and the latest ping seen at the loction during lunch hour. Call any such match a ``visit candidate''.  To get from visit candidates to visits, we impose the requirement that there be at least 2 pings in one of the location's geohash8s and that the dwelltime be at least 3 minutes. We also require that the visit be to a location that has no overlap with either the person's home geohash7 or the geohash7 we have identified as the person's narrow morning location so as to reduce the possibility of mis-identifying people living near a location or working at the location as visiting the location. In cases where a sequence of pings satisfying these criteria falls into the geohash8s of multiple locations we attribute the visit to the locations for which the dwelltime is longest.
% parameters checked by TS on 01/16

To put together our estimation dataset, we restrict the above visits to a set of users and restaurants we see sufficiently often. We require first that each user have at least 3 visits during the sample period, that each location have at least one visit by someone in the \emph{user base} per week on average, or at least five visits overall (from users overall, not just those in our \emph{user base}). This leaves us with 106,889 lunch visits by 9,188 users to 4,924 locations.
% number checked by TS on 01/14
% generated by source/analysis/descriptives/raw_data_descriptives.R

\begin{figure}
  \includegraphics[width=.8 \textwidth]{\figuredir/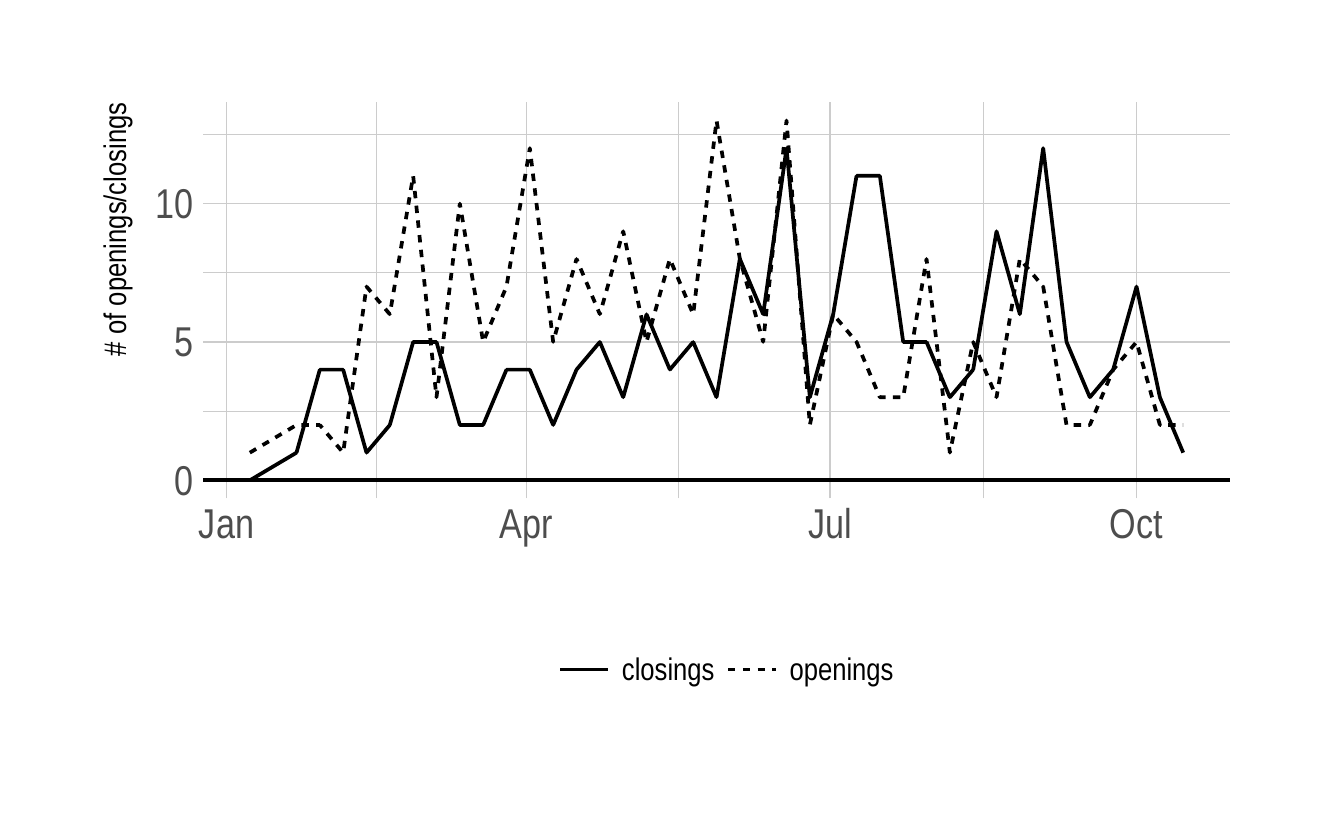}
  \caption{Restaurant Openings and Closings by Week}\label{fig:openclose}
\end{figure}

We also use data from Yelp to infer the dates of restaurant openings and closings.  We use the following heuristic: the opening is the date on which a listing was added to the Yelp database, while the closing date is the date on which a restaurant is marked by a member as closed.  Figure~\ref{fig:openclose} shows the openings and closings throughout the sample period.  We focus on openings and closings of restaurants that are considered by users whose morning location is within 3 miles of the opening/closing restaurant and who collectively take at least 500 lunch visits both before and after the change in status.
% numberes are correct, TS on 01/16

\paragraph{Distance}
As our measure of distance between a user's narrow morning location and each of the items in her choice set we use the simple straight-line distance (taking into account the earth's curvature). After calculating these distances we cull all alternatives that are further than 20 miles away from the choice set.
% parameter checked by TS on 01/16
% can be found in source/data/create_estimation_dataset.R

\paragraph{Item covariates}

The following restaurant covariates (or subsets thereof) are used in the estimation of both the MNL and the TTFM:

\begin{itemize}
  \setlength\itemsep{0em}
	\item \texttt{rating\_in\_sample}: the average rating awarded during the sample period Jan -- Oct 2017. If missing the value is replaced by the \texttt{rating\_in\_sample} average and another variable, \texttt{rating\_in\_sample\_missing} indicates that this replacement has been made
	\item \texttt{N\_ratings\_in\_sample}: the number of ratings that entered the computation of \texttt{rating\_in\_sample}
	\item \texttt{rating\_overall}: the average all--time rating. If missing the value is replaced by the \texttt{rating\_overall} average and another variable, \texttt{rating\_overall\_missing} indicates that this replacement has been made
	\item \texttt{N\_ratings\_overall}: the number of ratings that entered the computation of \texttt{rating\_overall}
	\item \texttt{category\_mexican} -- \texttt{category\_dancerestaurants}: A number of 0/1 indicator variables for whether an item has the corresponding category associate with it on Yelp
	\item \texttt{pricerange}: categorical variable indicating the restaurant's price category, from \texttt{\$} to \texttt{\$\$\$\$}
\end{itemize}
\afterpage{\clearpage}

\subsection{Estimation Details}\label{sec:estdetails}

To estimate the TTFM model, we build on the approach outlined in the appendix of \citet{ruiz2017shopper}, and indeed we use the same code base, since when we ignore the observable attributes of items, our model is a special case of \citeauthor{ruiz2017shopper}.  \citeauthor{ruiz2017shopper} considers a more complex setting where shoppers consider bundles of items. When restricted to the choice of a single item, the model is identical to TTFM replacing price with distance for TTFM.  However, we treat observable characteristics differently in TTFM than \citeauthor{ruiz2017shopper}.  In the latter, observables enter the consumer's mean utility directly, while in TTFM we incorporate observables by allowing them to shift the mean of the prior distribution of latent restaurant characteristics in a hierarchical model.  

We assume that one quarter of latent variables are affected by restaurant price range, one quarter are affected by restaurant categories, one quarter are affected by star ratings, and for one quarter of the latent variables there are no observables shifting the prior.  

The TTFM model defines a parameterized utility for each customer and restaurant,
\[U_{uit} = \underbrace{\lambda_i}_{\textrm{popularity}} + \underbrace{\theta_u^\top \alpha_i}_{\textrm{customer preferences}} - \underbrace{\gamma_u^\top \beta_i \cdot \text{log}(d_{uit})}_{\textrm{distance effect}} + \underbrace{\mu_i^\top \delta_{w_{ut}}}_{\textrm{time-varying effect}} + \underbrace{\epsilon_{uit}}_{\textrm{noise}},
\]
where $U_{uit}$ denotes the utility for the $t$-th visit of customer $u$ to restaurant $i$. This expression defines the utility as a function of latent variables which capture restaurant popularity, customer preferences, distance sensitivity, and time-varying effects (e.g., for holidays). All these factors are important because they shape the probabilities for each choice. Below we describe the latent variables in detail.

\noindent\textit{Restaurant popularity.} The term $\lambda_i$ is an intercept that captures overall (time-invariant) popularity for each restaurant $i$. Popular restaurant will have higher values of $\lambda_i$, which increases their choice probabilities.

\noindent\textit{Customer preferences.} Each customer $u$ has her own preferences, which we wish to infer from the data. We represent the customer preferences with a $k_1$-vector $\theta_u$ for each customer. Similarly, we represent the restaurant latent attributes with a vector $\alpha_i$ of the same length. For each choice, the inner product $\theta_u^\top \alpha_i$ represents how aligned the preferences of customer $u$ and the attributes of restaurant $i$ are. This term increases the utility (and consequently, the probability) of the types of restaurants that the customer tends to prefer.

\noindent\textit{Distance effects.} We next describe how we model the effect of the distance from the customer's morning location to each restaurant. We posit that each customer $u$ has an individualized distance sensitivity for each restaurant $i$, which is factorized as $\gamma_u^\top \beta_i$, where latent vectors $\gamma_u$ and $\beta_i$ have length $k_2$. Using a matrix factorization approach allows us to decompose the customer/restaurant distance sensitivity matrix into per-customer latent vectors $\gamma_u$ and per-restaurant latent vectors $\beta_i$, both of length $k_2$, therefore reducing the number of latent variables in the model. Thus, the inner product $\gamma_u^\top \beta_i$ indicates the distance sensitivity, which affects the utility through the term $-\gamma_u^\top \beta_i \cdot \text{log}(d_{uit})$. We place a minus sign in front of the distance effect terms to indicate that the utility decreases with distance.

\noindent\textit{Time-varying effects.} Taking into account time-varying effects allows us to explicitly model how the utilities of restaurants vary with the seasons or as a consequence of holidays. Towards that end we introduce the latent vectors $\mu_i$ and $\delta_w$ of length $k_3=5$. For each restaurant $i$ and calendar week $w$, the inner product $\mu_i^\top \delta_{w}$ captures the variation of the utility for that restaurant in that specific week. Note that each trip $t$ of customer $u$ is associated with its corresponding calendar week, $w_{ut}$.

\noindent\textit{Noise terms.} We place a Gumbel prior over the error (or noise) terms $\epsilon_{uit}$, which leads to a softmax model. That is, the probability that customer $u$ chooses restaurant $i$ in the $t$-th visit is
\[p(y_{ut}=i) \propto \exp\left\{ \lambda_i + \theta_u^\top \alpha_i - \gamma_u^\top \beta_i \cdot \text{log}(d_{uit}) + \mu_i^\top \delta_{w_{ut}} \right\},\]
where $y_{ut}$ denotes the choice.

\noindent\textbf{Hierarchical prior.} The resulting TTFM model is similar to the Shopper model \citep{ruiz2017shopper}, which is a model of market basket data. The TTFM is simpler because it does not consider bundles of products, i.e., we restrict the choices to one restaurant at a time, and thus we do not need to include additional restaurant interaction effects.

A key difference between Shopper and the TTFM is how we deal with low-frequency restaurants. To better capture the latent properties of low-frequency restaurants, we make use of observed restaurant attributes. In particular, we develop a hierarchical model to share statistical strength among the latent attribute vectors $\alpha_i$ and $\beta_i$.\footnote{We could also consider a hierarchical model over the time effect vectors $\mu_i$, but these are low-dimensional and factorize a smaller restaurant/week matrix, so for simplicity we assume independent priors over $\mu_i$.} Inspired by \citet{Zhao2017leveraging}, we place a prior that relates the latent attributes with the observed ones. More in detail, let $x_i$ be the vector of observed attributes for restaurant $i$, which has length $k_{\textrm{obs}}$. We consider a hierarchical Gaussian prior over the latent attributes $\alpha_i$ and distance coefficients $\beta_i$,
\begin{equation*}
	\begin{split}
		& p(\alpha_i \;|\; H_{\alpha}, x_i) = \frac{1}{(2\pi\sigma_\alpha^2)^{k_1/2}}\exp\left\{-\frac{1}{2\sigma_\alpha^2} ||\alpha_i-H_{\alpha}x_i||_2^2 \right\}, \\
		& p(\beta_i \;|\; H_{\beta}, x_i) = \frac{1}{(2\pi\sigma_\beta^2)^{k_2/2}}\exp\left\{-\frac{1}{2\sigma_\beta^2} ||\beta_i-H_{\beta}x_i||_2^2 \right\}.
	\end{split}
\end{equation*}
Here, we have introduced the latent matrices $H_{\alpha}$ and $H_{\beta}$, of sizes $k_1\times k_{\textrm{obs}}$ and $k_2\times k_{\textrm{obs}}$ respectively, which weigh the contribution of each observed attribute on the latent attributes. In this way, the (weighted) observed attributes of restaurant $i$ can shift the prior mean of the latent attributes. By learning the weighting matrices from the data, we can leverage the information from the observed attributes of high-frequency restaurants to estimate the latent attributes of low-frequency restaurants.

To reduce the number of entries of the weighting matrices, we set some blocks of these matrices to zero. In particular, we assume that one quarter of the latent variables is affected by restaurant price range only, one quarter is affected by restaurant categories, one quarter is affected by star ratings, and for the remaining quarter we assume that there are no observables shifting the prior (which is equivalent to independent priors). We found that this combination of independent and hierarchical priors over the latent variables works well in practice.

To complete the model specification, we place an independent Gaussian prior with zero mean over each latent variable in the model, including the weighting matrices $H_{\alpha}$ and $H_{\beta}$. We set the prior variance to one for most variables, except for $\gamma_u$ and $\beta_i$, for which the prior variance is $0.1$, and for $\delta_w$ and $\mu_i$, for which the prior variance is $0.01$. We also set the variance hyperparameters $\sigma_\alpha^2=\sigma_\beta^2=1$. 

\noindent\textbf{Inference.}
As in most Bayesian models the exact posterior over the latent variables is not available in closed form. Thus, we must use approximate Bayesian inference. In this work, we approximate the posterior over the latent variables using variational inference.

Variational inference approximates the posterior with a simpler and tractable distribution \citep{Jordan1999learn,Wainwright2008graphical}.
Let $\mathcal{H}$ be the vector of all hidden variables in the model, and $q(\mathcal{H})$ the variational distribution that approximates the posterior over $\mathcal{H}$. In variational inference, we specify a parameterized family of distributions $q(\mathcal{H})$, and then we choose the member of this family that is closest to the exact posterior, where closeness is measured in terms of the Kullback-Leibler (KL) divergence. Thus, variational inference casts inference as an optimization problem. Minimizing the KL divergence is equivalent to maximizing the evidence lower bound (ELBO),
\begin{equation*}
	\mathcal{L} = \mathbb{E}_{q(\mathcal{H})}\left[ \log p(y, \mathcal{H}) - \log q(\mathcal{H}) \right],
\end{equation*}
where $y$ denotes the observed data and $\mathcal{L}\leq \log p(y)$. Thus, in variational inference we first find the parameters of the approximating distribution that are closer to the exact posterior, and then we use the resulting distribution $q(\mathcal{H})$ as a proxy for the exact posterior, e.g., to approximate the posterior predictive distribution. For a review of variational inference, see \citet{Blei2017variational}.

Following other successful applications of variational inference, we consider mean-field variational inference, in which the variational distribution $q(\mathcal{H})$ factorizes across all latent variables. We use Gaussian variational factors for all the latent variables in the TTFM model, and therefore, we need to maximize the ELBO $\mathcal{L}$ with respect to the mean and variance parameters of these Gaussian distributions. We use gradient-based stochastic optimization \citep{Robbins1951stochastic,Blum1954approximation,Bottou2016optimization} to find these parameters. The stochasticity allows us to overcome two issues: the intractability of the expectations and the large size of the dataset. 

The first issue is that the expectations that define the ELBO are intractable. To address that, we take advantage of the fact that the gradient $\nabla\mathcal{L}$ itself can be expressed as an expectation, and we form and follow Monte Carlo estimators of the gradient in the optimization procedure. In particular, we use the reparameterization gradient \citep{Kingma2014autoencoding,Titsias2014doubly,Rezende2014stochastic}. The second issue is that the dataset is large. For that, we introduce a second layer of stochasticity in the optimization procedure by subsampling datapoints at each iteration and scaling the gradient estimate accordingly \citep{Hoffman2013stochastic}. Both approaches maintain the unbiasedness of the gradient estimator.

\subsection{Model Tuning and Goodness of Fit}\label{sec:fit}

\begin{figure}
  \begin{minipage}{\linewidth} \centering
    \includegraphics[height=.3 \textheight]{\figuredir/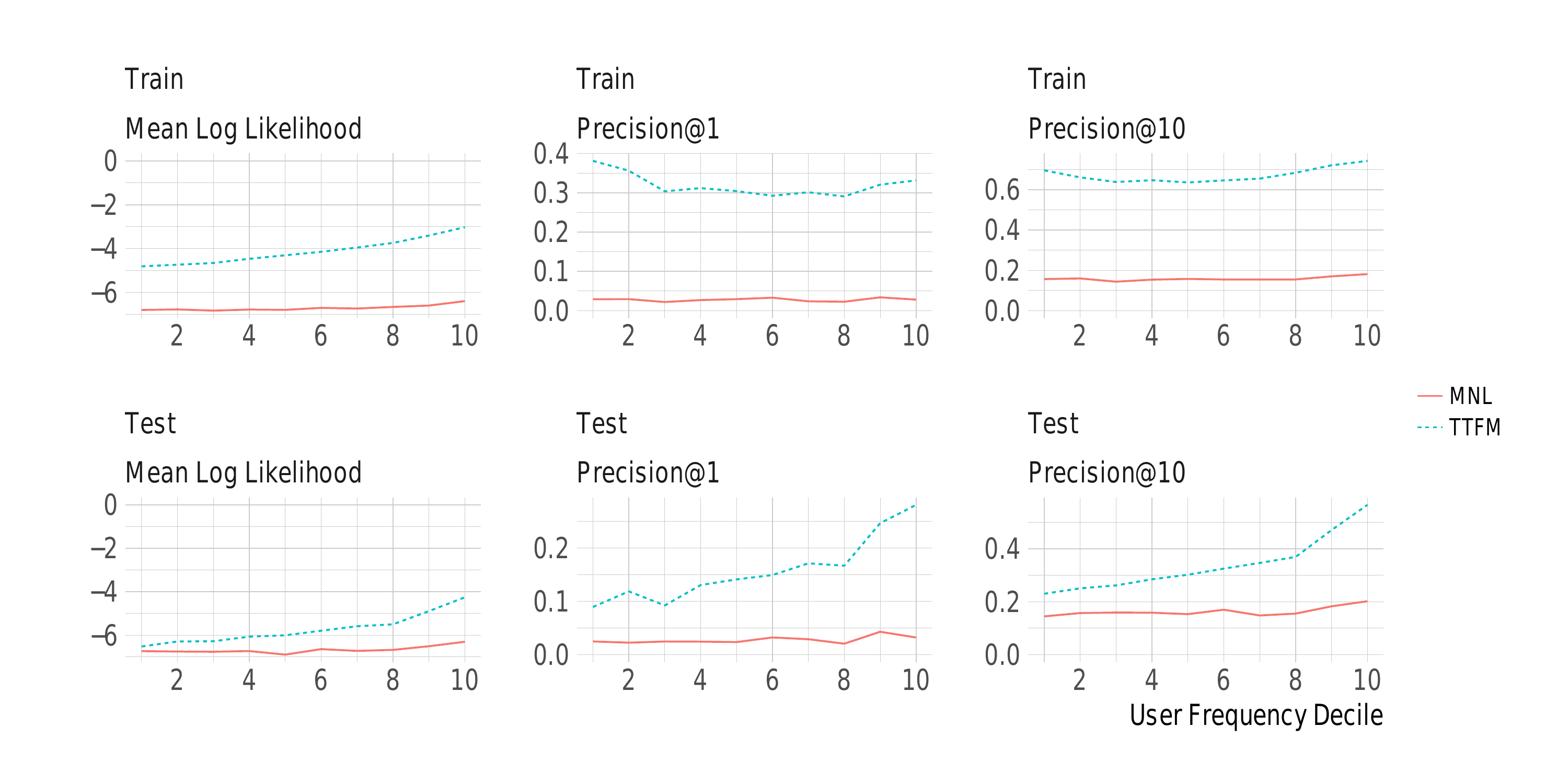}
    \caption{Goodness of Fit Measures by User Decile}  \label{fig:fitbyuserdecile}
  \end{minipage}
  \begin{minipage}{\linewidth} \centering
    \includegraphics[height=.3 \textheight]{\figuredir/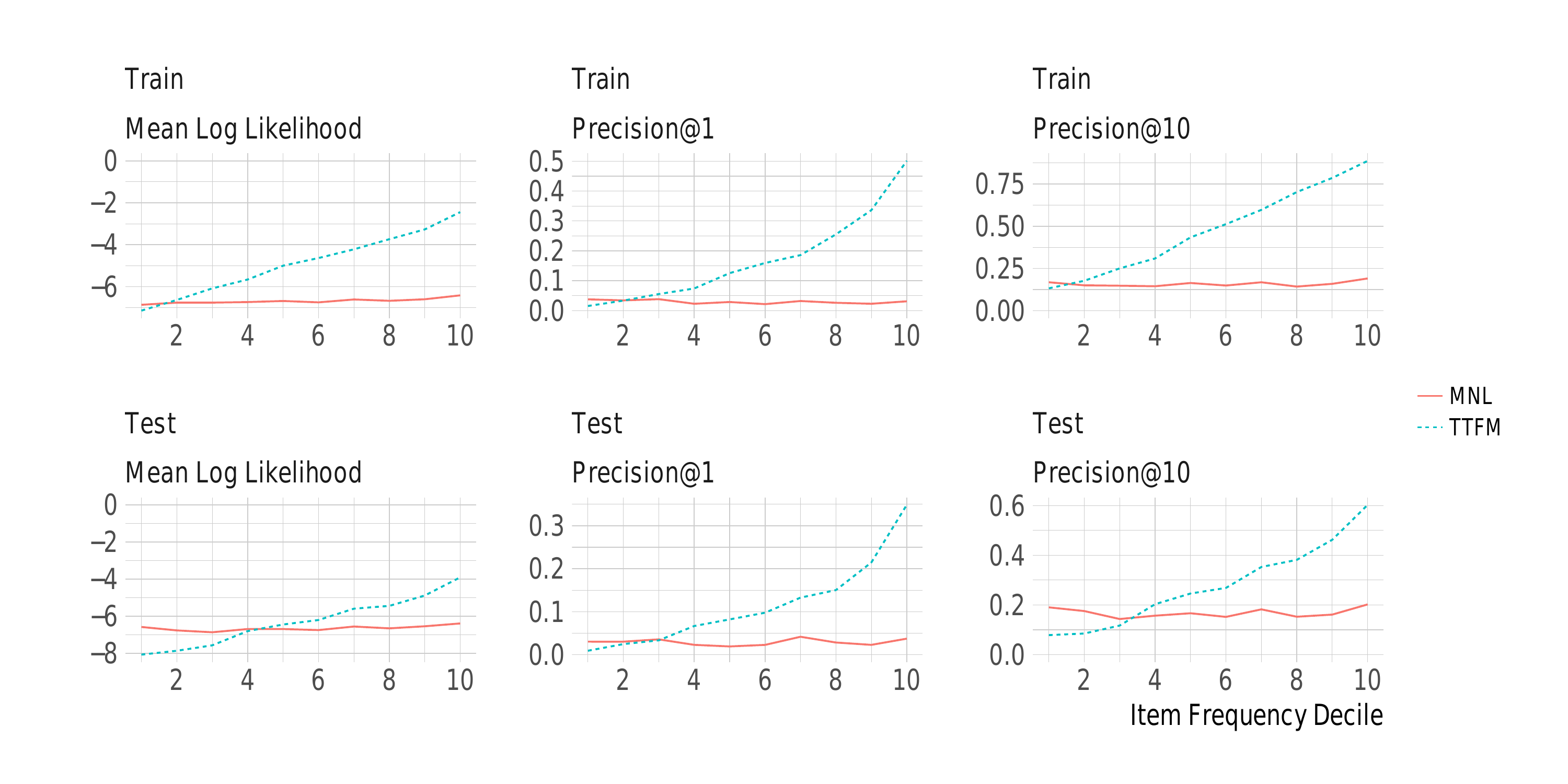}
    \caption{Goodness of Fit Measures by Restaurant Visit Decile}    \label{fig:fitbyrestdecile}
  \end{minipage}
  \begin{minipage}{\linewidth} \centering
    \includegraphics[height=.3 \textheight]{\figuredir/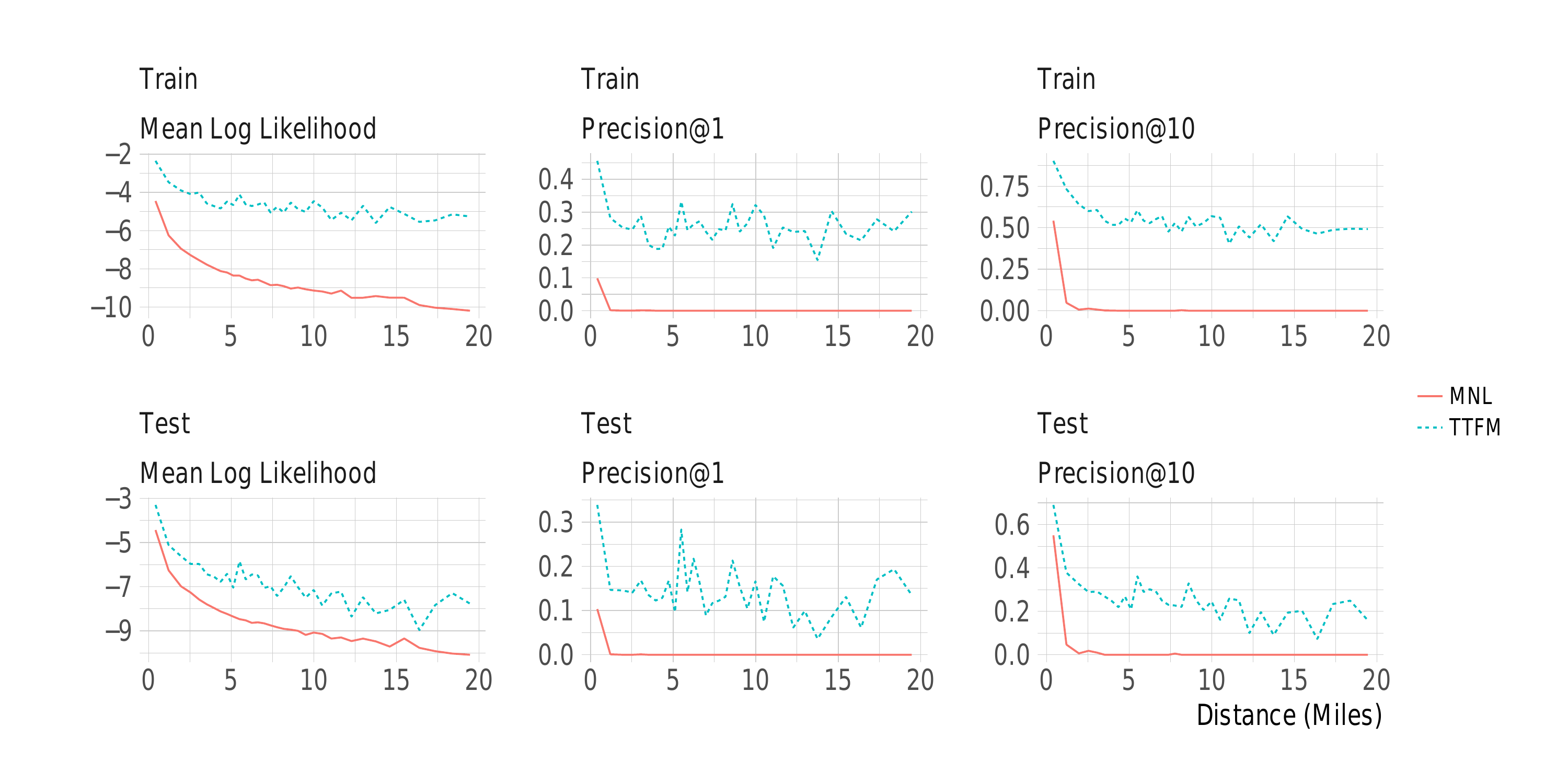}
    \caption{Goodness of Fit Measures by Distance}    \label{fig:fitbydistance}
  \end{minipage}
\end{figure}

 Figure~\ref{fig:demandbydistance} shows how well the model matches the actual purchase probabilities by distance.  Figures~\ref{fig:fitbyuserdecile},~\ref{fig:fitbyrestdecile} and~\ref{fig:fitbydistance} show goodness of fit broken out by distance from ther user, by user frequency decile, and by restaurant visit decile for the TTFM and MNL models.

\clearpage

\subsection{Additional Results}\label{sec:addlresults}

Table~\ref{tab:variancedecomp} illustrates how much of the variation in mean item utility (excluding distance) is explained by observable characteristics. All observables combined explain 14 percent of the variation. City and categories each explain 6 -- 7 percent and lose only a little explanatory power once other variables are accounted for. Star ratings and price range account for 2.8 and 2.3  percent of the variation respectively when considered alone, but only 0.6 percent and 0.4 percent once the other variables are taken into account.
% numbers checked by TS on 01/14
% generated by source/analysis/variance_decomposition.R

\begin{table}
  \caption{Contribution to Mean Item Utility of Observables}\label{tab:variancedecomp}
  \begin{tabular}{lrr}
  \toprule
Predictors & Variance contribution & Marginal variance contribution \\ 
  \midrule
Rating & 0.028 & 0.006 \\ 
  Price range & 0.023 & 0.004 \\ 
  City & 0.062 & 0.053 \\ 
  Categories & 0.067 & 0.046 \\ 
  All & 0.140 &  \\ 
   \bottomrule
\end{tabular}

\end{table}

\begin{table}[!h]
  \centering
  \caption{Distance Elasticities: Summary statistics}\label{tab:distance_elasticities}
   \small
    \begin{tabular}{lrrrrrr}
    \toprule
    Model             & \multicolumn{2}{c}{Overall} & \multicolumn{2}{c}{Within-User} & \multicolumn{2}{c}{Within-Item} \\
                      & Mean & SD                   & SD(Mean) & Mean(SD)      & SD(Mean) & Mean(SD) \\
    \cmidrule(r){1-1} \cmidrule(r){2-3}             \cmidrule(r){4-5}                 \cmidrule(r){6-7}
       TTFM & -1.4114 & 0.6810 & 0.5992 & 0.3005 & 0.2977 & 0.6003 \\ 
  MNL & -1.4291 & 0.0033 & 0.0001 & 0.0023 & 0.0002 & 0.0022 \\ 
  
      \bottomrule
    \end{tabular}
\end{table}

\afterpage{\clearpage}

\begin{figure}
  \includegraphics[width=.8 \textwidth]{\figuredir/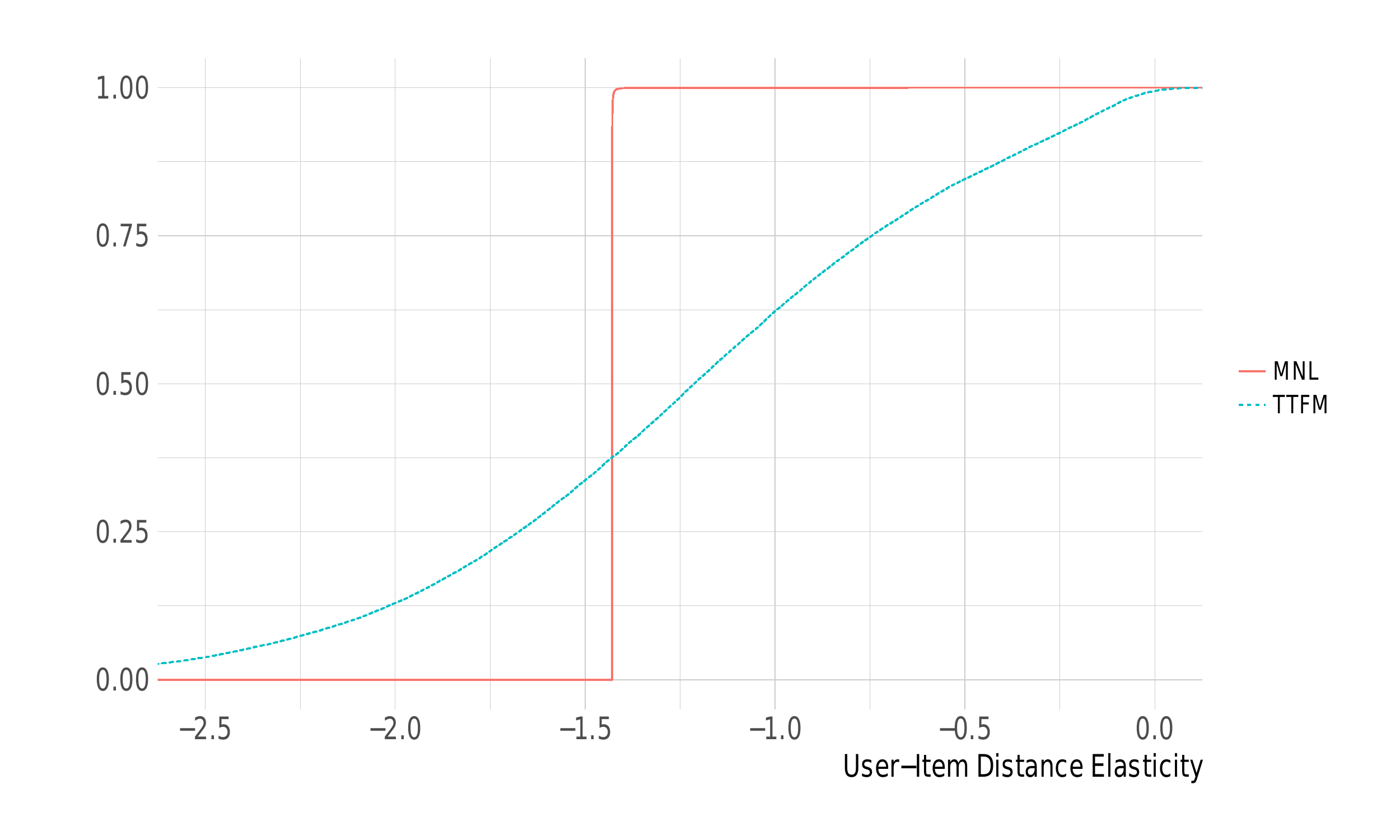}
  \caption{Distribution of Elasticities}\label{fig:elasticity}
\end{figure}

Table \ref{tab:distance_elasticities} gives the means and standard deviations of elasticities in the MNL and TTFM models. Figure~\ref{fig:elasticity} plots the distribution of elasticities where the unit of analysis is the restaurant-user pair.

\clearpage

%\begin{table}
 % \caption{Average Within-Item Elasticities by Restaurant Characteristics, MNL model.\label{tab:prefdistancebytype_mnl}}
 % \input{\tabledir/item_distance_elasticities_by_other_covariates_mnl.tex}
%\end{table}

%\begin{table}
 % \caption{Average Within-Item Elasticities by City, MNL model.}\label{tab:prefdistancebycity_mnl}
%  \input{\tabledir/item_distance_elasticities_by_city_mnl.tex}
%\end{table}

% \begin{figure}
% \caption{Average Within-Item Elasticities by City, TTFM model.}
% \label{fig:elasticityprefbyuser}
% \includegraphics[width=.8 \textwidth]{\figuredir/item_distance_elasticity_by_city_map.png}
% \end{figure}

%\subsection{Geohashes}\label{sec:geohashes}

%\begin{figure}
%\caption{geohash1s and 2s for the continental United States}
%\label{fig:geohash_illustration}
%\includegraphics[width=.8 \textwidth]{\figuredir/geohash_illustration.png}
%\begin{figurenotes}
%Source: \url{http://mapzen.github.io/leaflet-spatial-prefix-tree/}. Figure shows the two geohash1s --- \texttt{9} and \texttt{d} --- covering most of the continental United States, with geohash1 \texttt{d} being further subdivided into the 32 geohash2s from \texttt{90} to \texttt{9z}. 
%\end{figurenotes}
%\end{figure}
\afterpage{\clearpage}
\newpage

Tables \ref{tab:similar_locations_curry}, \ref{tab:similar_locations_chipotle} and \ref{tab:similar_locations_gofish} illustrate how the model can be used to
discover restaurants that are similar in terms of latent characteristics to a target restaurant.  Distance between two restaurants, $i$ and $i'$, 
is calculated as the Euclidean 
distance between the vectors of latent factors affecting mean utility, $\alpha_i$ and $\alpha_{i'}$. Note that because distance
is explicitly accounted for at the user level, we do not expect restaurants with similar latent characteristics to be near
one another; rather, they will uncover restaurants that would tend to be visited by the same consumers, if they
were (counterfactually) in the same location.  We see that indeed, the most similar restaurants to our target restaurants are in quite different geographic locations.  Perhaps surprisingly,
the category of the similar restaurants is generally different from the target restaurant, suggesting that other factors
are important to individuals selecting lunch restaurants.

\begin{table}
\caption{Locations similar (Latent Space) to Curry Up Now in Palo Alto (Indian Fast Food)}\label{tab:similar_locations_curry}
\centering
\resizebox{\linewidth}{!}{\begin{tabular}{lllrr}
\toprule
Location          & City           & Category     & Distance (Miles) & Latent Distance\\
\midrule
Zarzour Kabob \& Deli & San Jose   & Mideastern   & 17.2 & 1.58\\
Tava Kitchen      & Palo Alto      & Asian Fusion &  0.5 & 1.62\\
Pizza Hut         & Menlo Park     & Pizza        &  2.8 & 1.62\\
Subway            & Santa Clara    & Sandwiches   & 11.7 & 1.62\\
Rack \& Roll BBQ Shack & Redwood City & Seafood   &  3.8 & 1.62\\
\addlinespace
Burger King       & Redwood City   & Burgers      &  5.2 & 1.63\\
Subway            & Los Gatos      & Sandwiches   & 19.8 & 1.64\\
Pita Salt         & Campbell       & Street Food  & 17.0 & 1.64\\
Papa John's Pizza & San Jose       & Pizza        & 19.5 & 1.64\\
Cutesy Cupcakes   & San Jose       & Coffee       & 14.0 & 1.65\\
\bottomrule
\end{tabular}}

\end{table}

\begin{table}
\caption{Locations similar (Latent Space) to Chipotle Restaurant in Palo Alto (Tacos)}\label{tab:similar_locations_chipotle}
\centering
\resizebox{\linewidth}{!}{\begin{tabular}{lllrr}
\toprule
Location                     & City                & Category          & Distance (Miles) & Latent Distance\\
\midrule
The Van's Restaurant         & Belmont             & Sandwiches        &  8.5 & 1.28\\
La Viga Seafood Cocina Mexicana & Redwood City     & Mexican           &  4.1 & 1.31\\
Three Seasons                & Palo Alto           & Japanese          &  0.4 & 1.32\\
Cali Spartan Mexican Kitchen & San Jose            & Mexican           & 17.9 & 1.34\\
Poor House Bistro            & San Jose            & Southern          & 16.8 & 1.37\\
\addlinespace
McCormick Schmick's Seafood  & San Jose            & Trad American     & 17.3 & 1.38\\
Taqueria 3 Hermanos          & Mountain View       & Mexican           &  6.1 & 1.38\\
Peanuts Deluxe Cafe          & San Jose            & Breakfast         & 17.4 & 1.38\\
Izzy's San Carlos            & San Carlos          & New American      &  6.6 & 1.38\\
Bibo's Ny Pizza              & San Jose            & Pizza             & 18.0 & 1.39\\
\bottomrule
\end{tabular}}

\end{table}

\begin{table}
\caption{Locations similar (Latent Space) to Go Fish Poke Bar in Palo Alto}\label{tab:similar_locations_gofish}
\centering
\resizebox{\linewidth}{!}{\begin{tabular}{lllrr}
\toprule
Location          & City           & Category     & Distance (Miles) & Latent Distance\\
\midrule
Gourmet Franks                        & Palo Alto            & Hotdog         &  0.03  & 3.07\\
Lobster ShackXpress                   & Palo Alto            & Seafood        &  0.01  & 3.31\\
Mayfield Bakery \& Cafe               & Palo Alto            & New American   &  0.72  & 3.44\\
Shalala                               & Mountain View        & Japanese       &  6.15  & 3.46\\
Tin Pot Creamery                      & Palo Alto            & Coffee         &  0.70  & 3.47\\
\addlinespace
Mexican Fruit Stand                   & San Jose             & Street Food    & 18.63  & 3.60\\
Leonardo's Italian Deli \& Cafe       & Millbrae             & Coffee         & 16.50  & 3.62\\
Villa Del Sol Argentinian Restaurant  & South San Francisco  & Latin          & 19.84  & 3.63\\
Bobo Drinks Express                   & San Jose             & Coffee         & 19.34  & 3.63\\
Merlion Restaurant \& Bar             & Cupertino            & Bars           & 11.81  & 3.64\\
\bottomrule
\end{tabular}}

\end{table}
\newpage
\afterpage{\clearpage}

Tables \ref{tab:similar_locations_curry_util}, \ref{tab:similar_locations_chipotle_util} and \ref{tab:similar_locations_gofish_util} examine
restaurants that are similar accounting for all components of utility.  Let $U_{ui}$ be the average over dates $t$ that user $i$ visited restaurants
of $U_{uit}$. Distance between two restaurants, $i$ and $i'$, 
is calculated as the Euclidean 
distance between the mean utility vectors, $(U_{1i},..,U_{N_u,i})$ and $(U_{1i'},..,U_{N_u,i'})$, where $N_u$ is the number of users. Relative
to the previous exercise, we see that similar locations are very close geographically, but also still similar in other respects as well.  There are
many restaurants in close proximity to the selected restaurants, so the list displayed is {\emph not} simply the set of closest restaurants.

\begin{table}
\caption{Locations similar (Utility Space) to Curry Up Now in Palo Alto (Indian Fast Food)}\label{tab:similar_locations_curry_util}
\centering
\resizebox{\linewidth}{!}{\begin{tabular}{lllrr}
\toprule
Location          & City           & Category     & Distance (Miles) & Latent Distance\\
\midrule
Coupa Caf\'e           & Palo Alto  & Coffee       & 0.09 & 6.69\\
Cafe Venetia           & Palo Alto  & Coffee       & 0.14 & 7.54\\
Jamba Juice            & Palo Alto  & Juice        & 0.46 & 7.72\\
LYFE Kitchen           & Palo Alto  & New American & 0.17 & 7.74\\
Sancho's Taqueria      & Palo Alto  & Mexican      & 0.25 & 7.81\\
\addlinespace
T4                     & Palo Alto  & Coffee       & 0.18 & 7.89\\
Lemonade               & Palo Alto  & New American & 0.19 & 7.99\\
Coupa Caf\'e           & Palo Alto  & Coffee       & 0.28 & 8.17\\
Darbar Indian Cuisine  & Palo Alto  & Indpak       & 0.27 & 8.21\\
Gelataio               & Palo Alto  & Gelato       & 0.27 & 8.23\\
\bottomrule
\end{tabular}}

\end{table}

\begin{table}
\caption{Locations similar (Utility Space) to Chipotle Restaurant in Palo Alto (Mexican)}\label{tab:similar_locations_chipotle_util}
\centering
\resizebox{\linewidth}{!}{\begin{tabular}{lllrr}
\toprule
Location                      & City        & Category           & Distance (Miles)  & Latent Distance\\
\midrule
Bare Bowls                    & Palo Alto   & Juicebars          & 0.44 & 6.41\\
Coconuts Caribbean Restaurant & Palo Alto   & Caribbean          & 0.56 & 6.63\\
The Oasis                     & Menlo Park  & Bars               & 0.36 & 6.66\\
Coupa Caf\'e                  & Palo Alto   & Coffee             & 0.48 & 6.86\\
Pizza My Heart                & Palo Alto   & Pizza              & 0.44 & 7.07\\
\addlinespace
Fraiche                       & Palo Alto   & Coffee             & 0.48 & 7.21\\
Cafe Del Sol Restaurant       & Menlo Park  & Mexican            & 0.86 & 7.23\\
MP Mongolian BBQ              & Menlo Park  & BBQ                & 0.68 & 7.34\\
Bistro Maxine                 & Palo Alto   & Breakfast          & 0.49 & 7.85\\
Koma Sushi Restaurant         & Menlo Park  & Japanese           & 0.35 & 7.88\\
\bottomrule
\end{tabular}}

\end{table}

\begin{table}
\caption{Locations similar (Utility Space) to Go Fish Poke Bar in Palo Alto}\label{tab:similar_locations_gofish_util}
\centering
\resizebox{\linewidth}{!}{\begin{tabular}{lllrr}
\toprule
Location                   & City        & Category      & Distance (Miles)  & Latent Distance\\
\midrule
Crepevine Restaurant       & Palo Alto   & New American  & 0.61 & 17.96\\
California Pizza Kitchen   & Palo Alto   & New American  & 0.10 & 18.03\\
True Food Kitchen          & Palo Alto   & New American  & 0.08 & 19.67\\
Joya Restaurant            & Palo Alto   & Mexican       & 0.58 & 19.85\\
Gott's Roadside            & Palo Alto   & Bars          & 0.68 & 20.18\\
\addlinespace
Pressed Juicery            & Palo Alto   & Juice         & 0.03 & 20.37\\
American Girl              & Palo Alto   & Trad American & 0.10 & 20.37\\
Dashi Japanese Restaurant  & Menlo Park  & Japanese      & 2.75 & 20.78\\
Cafe Bistro                & Palo Alto   & New American  & 0.30 & 20.84\\
NOLA Restaurant            & Palo Alto   & Bars          & 0.54 & 21.01\\
\bottomrule
\end{tabular}}

\end{table}
\newpage
\afterpage{\clearpage}

\subsection{Counterfactual Calculations}\label{sec:counterfactuals}

\begin{figure}
  \includegraphics[width=.8 \textwidth]{\figuredir/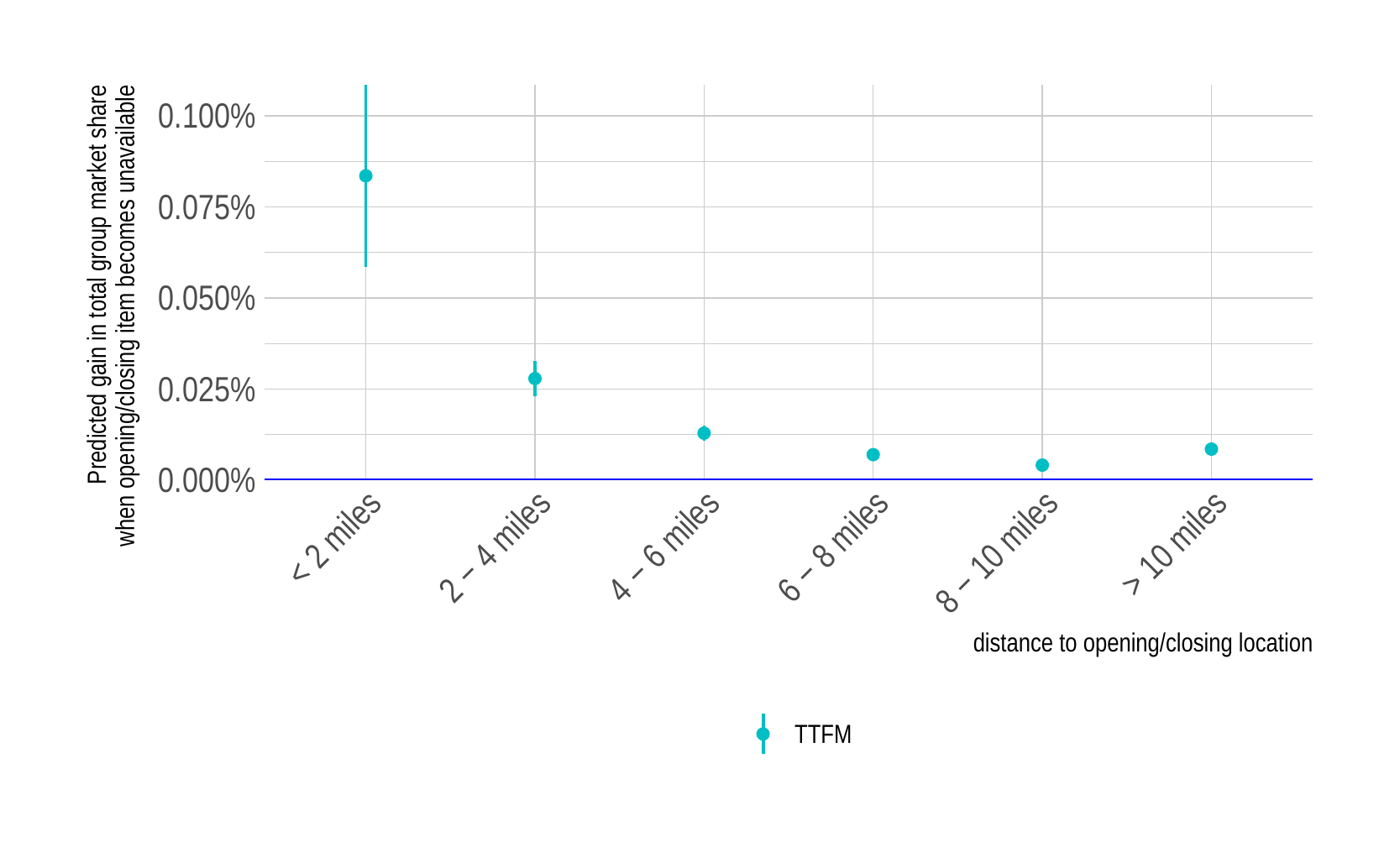}
  \caption{Model Predictions of the Effect of Restaurant Openings and Closings Controlling for Other Changes.}\label{fig:OPCL1}
  \begin{figurenotes}
  The figure shows the average of the predicted difference in the total market share of each group between the period where the target restaurant is closed and when it is open, minus the difference between the two periods predicted by the model in the counterfactual scenario where the target restaurant is closed in both periods. The user base for the calculated market shares includes all users from the full sample whose morning location is within three miles of the target restaurant and who visit at least one restaurant in both periods. User-item market shares under each regime (target restaurant open and target restaurant closed) are averaged using weights proportional to each user's share of visits in the group to any location during the open period. The bars in the figure show the point estimates plus or minus two times the standard error of the estimate, which is calculated as the standard deviation of the estimates across the different opening or closing events divided by the square root of the number of events.
  \end{figurenotes}
\end{figure}

Figure~\ref{fig:OPCL1} illustrates the model's predicted impact of restaurant openings and closings on different groups of neighboring restaurants.

Sections \ref{sec:marketfit} and the counterfactual exercise in
\ref{sec:openclose} rely on a similar form of calculation: how many visits would
we predict restaurant $i'$ would receive if it were located in location currently
occupied by restaurant $i$. When we do this, we assume that all characteristics
of $i'$, both observed and latent stay the same, except that when we calculate the utility
for each consumer for $i'$, we use the location of $i$ when calculating distances.
In principle, we can predict the demand
$i'$ would receive at any location in the region, however it is easier to have
$i'$ replace an existing location $i$, since this ensures that the chosen
location is reasonable (e.g. not in the middle of a forest or a highway).

To calculate demand for $i'$ replacing restaurant $i$, we calculate new values of
the utilities for $i'$ for each user $u$ and session $t$, which change only due
to the new distances $d_{ui}$ are used instead of the real distances $d_{u,i'}$.
\[ U_{uti';i} = U_{uti'} - \gamma_u \beta_{i'} \left( \log(d_{ui}) - \log(d_{ui'}) \right)\]
Then we recalculate each user's new choice probabilities in each session, and
take the sum across all users and sessions in order to get the new predicted
total demand for each restaurant under the counterfactual that $i'$ is located in
the location of restaurant $i$.
\[ P(y_{uti';i} = 1) = \frac{\exp(U_{uti';i})}{\exp(U_{uti';i}) + \sum_{l \notin {i,i'}} \exp(U_{utl}) } \]
\[ \text{Demand}_{i';i} = \sum_u \sum_t P(y_{uti';i} = 1) \]

In Section \ref{sec:openclose}, we repeat this calculation for each restaurant $i$ that either opens or closes. We draw $i'$ from two distinct sets, $I_{same}$ is 100
restaurants chosen at random from the same category as $i$ and $I_{diff}$
is 100 restaurants chosen at random from restaurants that are not in the same
category as $i$. In Table \ref{tab:CF} we compare the predicted demand for the
place that opens or closes,  $\text{Demand}_{i;i}$, 
to the mean counterfactual
predictions for $i'$ in $I_{same}$ and $I_{diff}$, i.e., 
\[\frac{1}{|I_{same}|} \sum_{i' \in I_{same}} \text{Demand}_{i';i}\]
\[\frac{1}{|I_{diff}|} \sum_{i' \in I_{diff}} \text{Demand}_{i';i}\]

\begin{figure}

  \centering
  \begin{minipage}{0.45\textwidth} \centering
    \includegraphics[width=\textwidth]{\figuredir/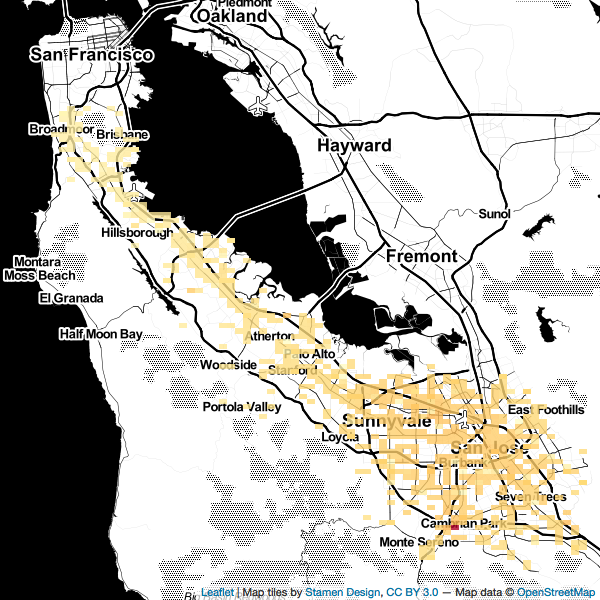}
    Cafes
  \end{minipage}
  \begin{minipage}{0.45\textwidth} \centering
    \includegraphics[width=\textwidth]{\figuredir/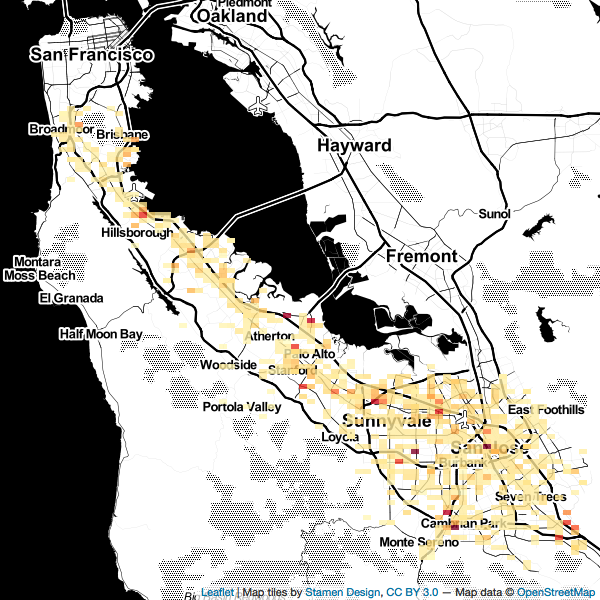}
    Chicken Wings
  \end{minipage}

  \begin{minipage}{0.45\textwidth} \centering
    \includegraphics[width=\textwidth]{\figuredir/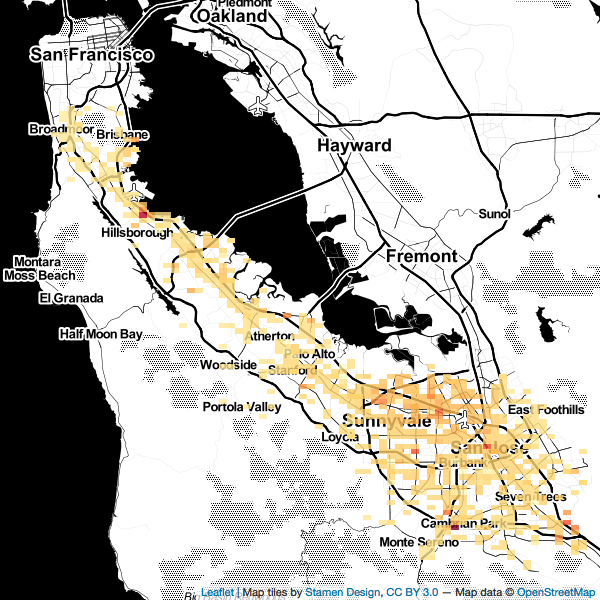}
    Filipino Restaurants
  \end{minipage}
  \begin{minipage}{0.45\textwidth} \centering
    \includegraphics[width=\textwidth]{\figuredir/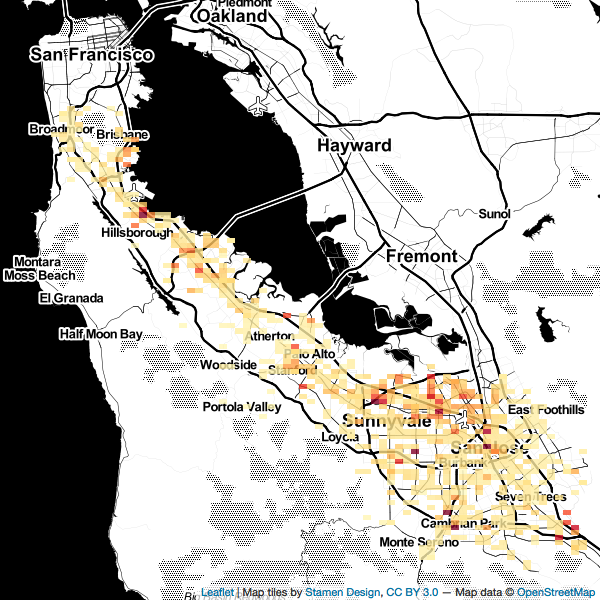}
    Sandwiches
  \end{minipage}

  \begin{minipage}{0.45\textwidth} \centering
    \includegraphics[width=\textwidth]{\figuredir/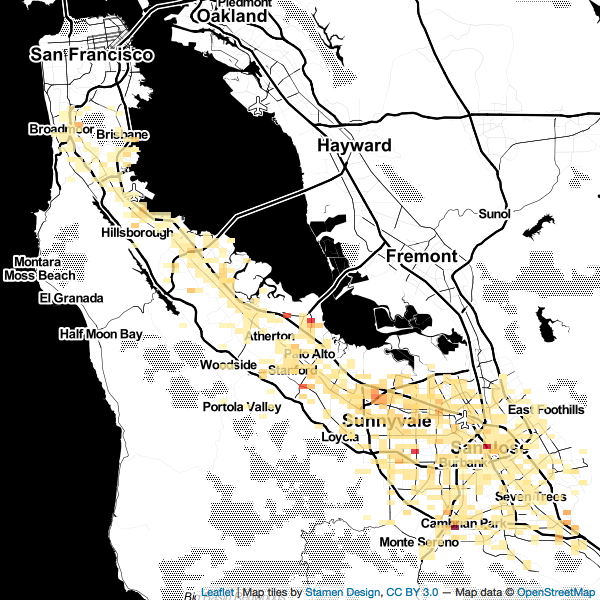}
    Vegetarian
  \end{minipage}
  \begin{minipage}{0.45\textwidth} \centering
    \includegraphics[width=\textwidth]{\figuredir/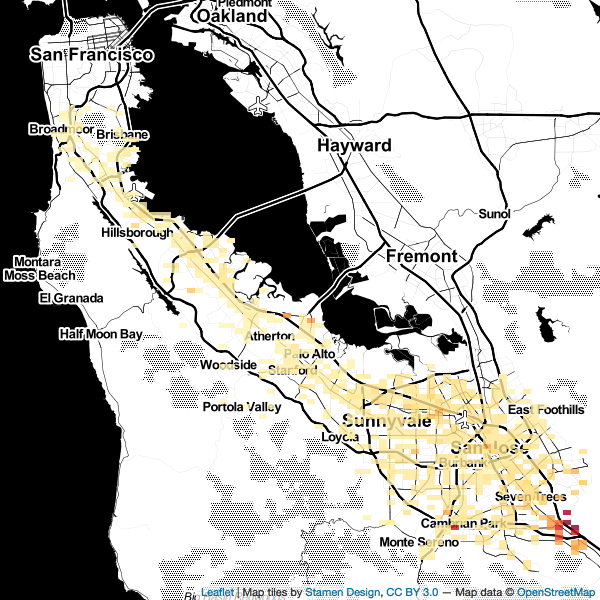}
    Vietnamese Restaurants
  \end{minipage}

  \caption{Best Locations for Restaurant Category}  \label{fig:bestlocation}
\end{figure}

\begin{figure}

  \centering
  \begin{minipage}{0.45\textwidth} \centering
    \includegraphics[width=\textwidth]{\figuredir/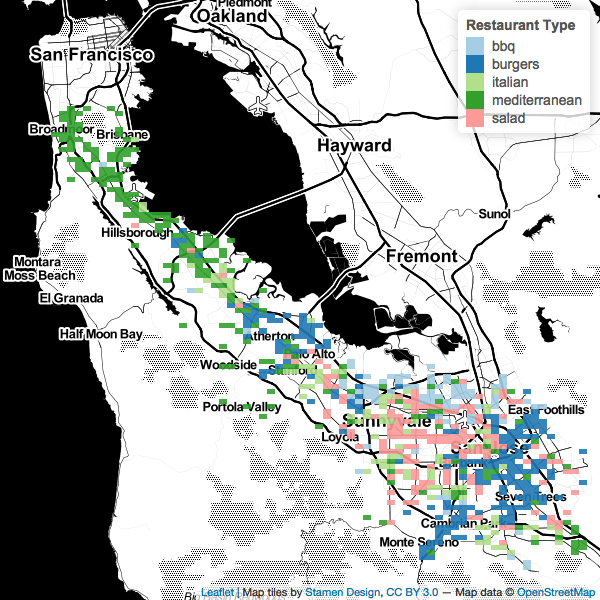}
    Mid-Priced (\$\$) Western Cuisine
  \end{minipage}
  \begin{minipage}{0.45\textwidth} \centering
    \includegraphics[width=\textwidth]{\figuredir/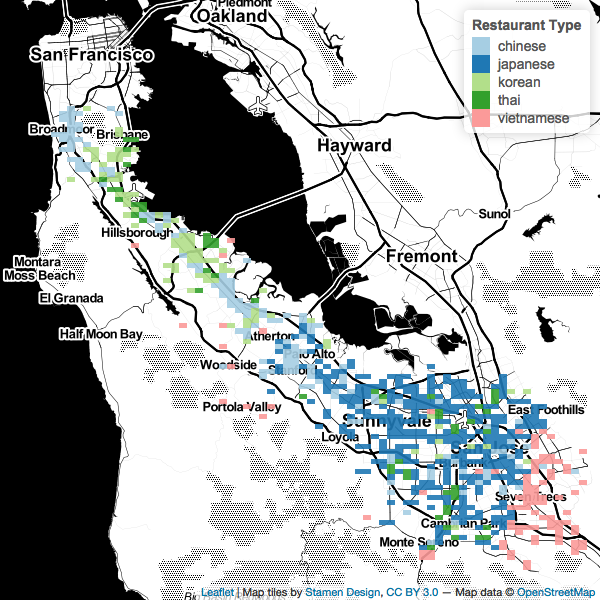}
    Mid-Priced (\$\$) Asian Cuisine
  \end{minipage}

  \begin{minipage}{0.45\textwidth} \centering
    \includegraphics[width=\textwidth]{\figuredir/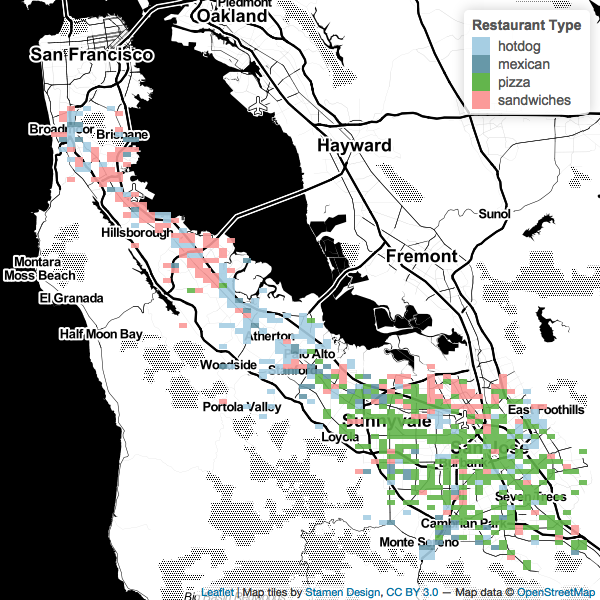}
    Cheap (\$) Fast Food
  \end{minipage}
  \begin{minipage}{0.45\textwidth} \centering
    \includegraphics[width=\textwidth]{\figuredir/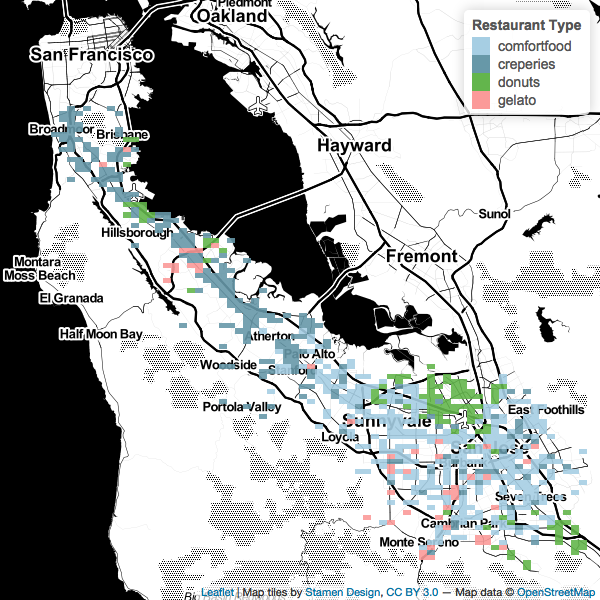}
    Cheap (\$) Treats
  \end{minipage}

  \caption{Best Restaurant Category for Locations} \label{fig:bestcategory}
\end{figure}

In Section \ref{sec:marketfit}, the set of target restaurants includes one location selected at random
from each geohash6. The set $I_{alt}$ is one restaurant from each major category (the
variable \verb!category_most_common!) with the constraint that each restaurant
chosen is within $0.1$ standard deviation of the population mean for total
demand. This constraint was to try to make the set of comparison restaurants
relatively similar in popularity. In the ``best location for each category'' in
Figure \ref{fig:bestlocation} we plot for a single category $i'$ the predicted demand
$\text{Demand}_{i';i}$ for each $i$ in the set of target locations. 
In Figure \ref{fig:bestcategory}, we selected subsets of 4
or 5 categories of restaurants from $I_{alt}$ that have the same price range and illustrate for each target location the category of restaurant that is
$\text{argmax}_{i' \in I_{alt}} \text{Demand}_{i';i}.$ 

\end{document}